\documentclass[twocolumn,aps,showpacs,prc]{revtex4}
\usepackage{graphicx}
\usepackage{longtable}
\usepackage{dcolumn}
\usepackage{bm}
\usepackage{multirow}

\begin{document}

\title{
      Fission half-lives of super-heavy nuclei in a microscopic approach}
\author{M. Warda}%
\email{warda@kft.umcs.lublin.pl}
\affiliation{{\it Katedra Fizyki Teoretycznej, Uniwersytet 
Marii Curie-Sk\l odowskiej, ul. Radziszewskiego 10, 20-031 Lublin, Poland}}
\author{J.L. Egido}%
\email{j.luis.egido@uam.es}%
\affiliation{%
{\it Departamento de F\'{\i}sica Te\'orica, 
Universidad Aut\'onoma de Madrid, 28049 Madrid, Spain}}%


\date{\today}

\begin{abstract}

A systematic study of 160 heavy and super-heavy nuclei is performed  in the
Hartree-Fock-Bogoliubov approach with the finite range and density dependent Gogny force with the D1S parameter set. 
We show calculations in several approximations:  with axially symmetric and reflexion symmetric
wave functions, with axially symmetric and non-reflexion symmetric wave functions and finally some
representative examples with triaxial wave functions are also discussed.

Relevant properties of the ground state and  along the fission path  are thoroughly analyzed.
Fission barriers, Q$_\alpha$-factors and lifetimes with respect to fission and $\alpha$-decay as well as other
observables are discussed. Larger configuration spaces and more general HFB wave functions as compared to  previous studies provide a very good agreement with the experimental data.
\\   \\
\end{abstract}
\maketitle

\section{Introduction}       

The stability and structure of nuclei at the upper end of the nuclear chart is a  hot topic in contemporary nuclear physics. 
It is a challenging task to answer  which heavy nuclides may exist and what properties they may have. Therefore                                              
strong efforts have been made in the experimental  developments as well as in the theoretical description of super-heavy elements (SHE's).

In the last decades a huge progress in the synthesis of new elements has been achieved in world leading laboratories like the GSI, Darmstadt \cite{Hof89,Goe99,hof07,hes09,dul10,hes10,khu10,gat11}, JINR, Dubna \cite{oga05,oga07,oga07a,oga09,oga10,oga12}, and  RIKEN, Tokyo \cite{mor04,mor07,mor09,hab11}. The first SHE's  with $Z \le 113$ and $N \le 165$ were produced in  cold fusion reactions. In these experiments, involving neutron rich projectiles and spherical targets ($^{208}$Pb or $^{209}$Bi), weakly excited compound nuclei were produced which cooled down by the emission of only one or two neutrons. Further experimental progress in the synthesis of the heaviest elements was achieved  by hot fusion reactions in which targets of deformed actinide    were bombarded with the doubly-magic nucleus $^{48}$Ca. The compound nucleus created in this way was more  excited and three or more neutrons were emitted. These reactions succeeded in the synthesis of new elements up to  $Z= 118$ and $N=176$ \cite{oga07}. The first observation of the element 117 was possible lately \cite{oga10} using a radioactive $^{249}$Bk target. New possibilities for the synthesis of new isotopes will be  opened  in heavy ion collisions with radioactive ion beams \cite{sos09,ger09}.  
Nowadays, many other laboratories are involved in the exploration of SHE's. Thus the experimental groups from Berkeley \cite{sta09}, GANIL \cite{sos09},  Livermore \cite{mod04,oga07a}, Jyv\"askyl\"a \cite{the07,jul10}, and  Oak Ridge \cite{oga12} are working in this direction. They would bring in the nearest future further information on the stability and properties of the SHE's and an independent verification of  the  existing data.

In a parallel way to the experimental efforts, the properties of the SHE's have been also investigated in various nuclear models.  A proper description of trans-fermium nuclei is a great challenge for any theoretical  model. Usually,  the parameters of the  theories on atomic nuclei are 
 adjusted to the stable isotopes and then extrapolated  to the region of heavier systems.
 Therefore many tries in different theoretical approaches are performed to foresee the stability and the structure of the heaviest nuclei. A detailed review of the theoretical analysis of SHE's can be found in Ref. \cite{sob07}.
The first theoretical investigations on the stability of heavy nuclei were made in the 60's. It was noticed that shell effects could  stabilize nuclei heavier than those known at that time \cite{sob66,mye66}. Calculations made in the macroscopic-microscopic model with Strutinsky shell correction predicted the ``island of stability". Large values of shell energies  were obtained at $Z=108, N=162$ for prolate deformed nuclei and  $Z=114, N=184$ for spherical ones \cite{sob66, mos69}. 
In the last decades many calculations have been made providing  more and more precise predictions. The fission barriers and the ground state properties were calculated using macroscopic-microscopic methods with large range of deformation parameters and nuclear shapes, including reflection and axial symmetry breaking \cite{bar81,loj85,sta99,mun03,kow10,Moe87,Moe89,Moe00,Moe01,Moe92,gam05,kum12}.

Self-consistent methods also provided many results on  fission barriers and half-lives. Important results have been obtained in the relativistic mean-field (RMF) \cite{bur04,Ben98,kar10,kum12}, the Hartree-Fock (HF) approach with Skyrme forces \cite{flo75,Cwi96,Rut97,Ben98,ben99,bur04,sta05}  and the Hartree Fock-Bogoliubov (HFB) theory with Gogny forces \cite{ber00,ber01,war06}. The first calculations of fission barriers were performed in the axial and reflection symmetric regime but later all relevant deformations were considered in the minimization of the energy.

It is well known that the liquid drop model does not predict a fission barrier in the heaviest nuclei and that the stability of the trans-fermium nuclei is achieved by the  shell effects. The self-consistent quantum mechanical methods (RMF, HF and HFB) as  microscopic theories are the perfect tools for the analysis of SHE's. Moreover, in the self-consistent calculations all possible shapes of a nucleus are considered in the minimization process. In contrast, the most of the macroscopic-microscopic models are restricted to some pre-defined classes of deformations and only ``optimal shapes" \cite{iva09} allow to obtain any configuration of a nucleus. Therefore they are very suitable to describe large deformations of nuclei around the scission point.
  A degree of freedom which plays an important role along the fission path are the pairing correlations \cite{sta89}. Since in the HFB theory the particle-hole and the particle-particle  matrix elements are treated on the same footing,  the proper consideration of  pairing  along the whole fission path is guaranteed. This method has been successfully applied in many aspects of low energy nuclear physics, in particular  in the description of fission barriers of heavy nuclei \cite{ber00,ber01,werp}. 
 Another theoretical quest was  to discover a semi-empirical formula describing  $\alpha$ emission half-lives \cite{den10,das09}. These investigations are very important as  $\alpha$ radioactivity is the dominant decay channel in many SHE's.

The purpose of this article is to perform a systematic study of SHE's with respect to their  stability and ground state properties
in the framework of the HFB theory with  the density dependent finite range Gogny  force and the D1S parametrization. In our analysis we include the region of the well known Fermium (Fm, $Z= 100$) and Nobelium (No, $Z= 102$)  elements to compare our predictions with the available experimental data. We show results for the heavier even-even nuclei Rutherfordium  (Rf, $Z= 104$), Seaborgium (Sg, $Z= 106$), Hassium (Hs, $Z= 108$), Darmstadtium (Ds, $Z= 110$) and Copernicium (Cn, $Z= 112$, which was named  two years ago \cite{iupac}). Heavier elements with $Z= 114-124$, without given name so far, are also considered. We limit our study to $N\le190$ isotopes. These nuclei are nowadays in the main stream of interest of experiments with SHE's. 

A large amount of information on nuclear structure and stability can be obtained from properties of the ground states of the SHE's. Consequently, we  start our investigation with the description of the ground state characteristics. Deformations, pairing energies and two-nucleon separation energies are analyzed and collated with the single-particle energy level scheme.  The ground state energies can be used to evaluate the  $Q_\alpha$ values which are necessary to calculate the probability of $\alpha$ emission - one of the dominant decay modes in SHE.
A competing process to $\alpha$ decay is the spontaneous fission. To analyze this mode we determine the fission barriers for all mentioned SHE's as a function of the quadrupole moment $Q_2$. The calculations were performed in an axial basis, although we are aware of the non-axial effects on the height of the barrier and we discuss them in a few selected cases. The impact  of the octupole deformation on the potential energy along the fission path is crucial in the determination of fission barriers of SHE's. This  is taken into account by allowing non-reflexion symmetrical shapes. The calculations were performed in a large deformed harmonic oscillator basis paying special attention to the proper optimization of the oscillator lengths and to the convergence of the calculations with the size of the basis. Next, using the WKB approximation, we calculate the fission half-lives. Finally, the comparison of the half-lives for
  $\alpha$ decay and spontaneous fission allows to predict the stability of the heaviest nuclei. 

This article is organized as follows. In Sec. II we briefly describe the constrained Hartree-Fock-Bogoliubov calculations. The description of the ground-state properties of SHE's is shown in Sec. III,  fission barriers in Sec. IV. and  half-lives of SHE's are discussed in Sec. V. Finally, Sec. VI contains a summary and some concluding remarks.

\section{Theoretical model}

  In our research we will apply the self consistent Hartree-Fock-Bogoliubov theory with the finite range density dependent Gogny force. 
 In the numerical applications
we use the D1S \cite{Ber84,Ber91} parameterization of the Gogny interaction. The D1S
parameters were adjusted \cite{Ber84} to give a better surface energy term (crucial for
a proper description of the fission phenomenon). 
The choice of the Gogny force with the D1S parameterization is based on the fact
that whenever this interaction has been used to describe low energy nuclear
structure phenomena an, at least, reasonable agreement with experiment has 
been obtained. This degree of agreement has been obtained both for calculations 
at the mean field level and beyond
\cite{Dec80,Ber89,Ber96,gir83,egi89,gir89,egi93,bla95,gar98,Rod02,war05,war05a,war06a,Rod07,Rod10,Lop11,war11}.


\subsection{Details of self-consistent HFB calculations}
In the microscopic HFB calculations we have used the computer code presented in
Ref.~\cite{Egi97}, see also \cite{werp} where special attention was paid to an accurate computation of the
matrix elements of the Gogny interaction for very large basis like the one used
in this paper.
The self-consistent equations have been
solved  by expanding  the  quasiparticle creation and annihilation
operators on  finite bases of axially symmetric deformed harmonic oscillator
(HO) eigenfunctions.  The size of the bases used depends upon two parameters,
$N_0$ and $q$, which are  related to the allowed range of the HO quantum
numbers through the relation
$$
\frac{1}{q} n_z + (2n_\perp + |m|) \le N_0.
$$
Along the perpendicular direction we take $N_0$ shells, (i.e. $2n_\perp +
|m|=0,\ldots,N_0$) and along the $z$ direction we include up to $qN_0$ shells
depending on the value of $2n_\perp +|m|$. In the present study we have used
$q=1.5$, a value which is suited for the elongated shapes along the $z$
direction typical of the fission process, and $N_0=15$.  Other parameters characterizing
 the HO bases are the
oscillator lengths $b_\perp$ and $b_z$. These two quantities have been
determined, for each calculated wave function, as to minimize the HFB energy.
In order to study the different paths to fission,  in our
calculations we have used  as constraints  the axial quadrupole ($Q_2$) and octupole
($Q_3$)  moments, with $ \hat Q_\lambda=r^\lambda P_\lambda (\cos (\theta))$.
Higher multipolarities are adjusted in the self-consistent process to minimize the energy.
To study the impact of triaxiality effects  we have also 
carried out calculations for  a few nuclei where the axial symmetry requirement was released but the left-right symmetry was
imposed. 

In the calculations the Coulomb exchange energy has been treated in the Slater approximation 
\cite{tit74,Ang01}. 
 Additionally,  the Coulomb and the spin-orbit contributions  to the pairing field have been neglected.
Finally, the two body kinetic energy correction (2b-KEC) is not included in the variation
process because, for heavy nuclei, it remains almost constant for most of the
physical configurations. As this  term was included in the fitting of the
force, we have to include its contribution at the end of the calculation in
order to obtain reasonable binding energies. See also \cite{werp,war11} for a detailed discussion of
the relevance of neglected or approximated terms in these calculations.

We have also subtracted from the HFB energy the  rotational energy
corrections  (REC) stemming from the restoration of the rotational symmetry. This
correction has a considerable influence on the energy landscape (and therefore
on the height of the fission barriers) as it is somehow proportional to the degree
of symmetry breaking and therefore proportional to the quadrupole moment. A
full calculation of the REC would imply an angular momentum projection
\cite{Rod02,Rod01,die64} which is only feasible for light nuclei. In order to
estimate the REC we have followed the usual recipe \cite{Ring80} of
subtracting to the HFB energy the quantity  $ \langle \Delta
\vec{J}^{2}\rangle /(2{\mathcal{J}}_{Y}) $, where  $ \langle \Delta
\vec{J}^{2}\rangle $ is the fluctuation  associated with the angular momentum 
operators in the HFB wave function and $ {\mathcal{J}}_{Y} $  is the Yoccoz moment of
inertia \cite{Ring80}. This moment of inertia has been computed using the
``cranking" approximation in which the full linear  response matrix appearing
in its expression is replaced by the zero order  approximation. The effect of
the ``cranking approximation" in the Yoccoz moment of inertia was analyzed with
the Gogny interaction for heavy nuclei in \cite{Egi00} by comparing it  with
the one extracted from an angular momentum projected calculation (see also
\cite{Rod01} for a comparison in light nuclei). The conclusion is that the
exact REC is a factor 0.7 smaller than the one computed with the ``cranking"
approximation to the Yoccoz moment of inertia for strongly deformed
configurations (a similar behavior has been observed for the Thouless-Valatin
moment of inertia in \cite{Gir92}). We have taken this phenomenological factor
into account in our calculation of the REC.


\subsection{Evaluation of lifetimes}
The evaluation of the spontaneous fission half-life is carried out in the standard WKB framework where
$T_{\rm sf}$ is given (in seconds) by \cite{hil53}
\begin{equation}
T_{\rm sf} = 2.86 \cdot 10^{-21} (1+\exp (2S)) \,\,.
\label{TSF}
\end{equation}
In this expression $S$ is the action along the $Q_2$ constrained path which is
given by 
\begin{equation}
S = \int_a^b dQ_2 \sqrt{2B(Q_2)(V(Q_2)-E_0)}\,\,.
\label{ACTION}
\end{equation}
$a$ corresponds to the $Q_2$ value of the ground state and $b$ the $Q_2$ value where the
potential energy equals the one of the ground state. 
For the collective quadrupole inertia $B(Q_2)$ we have used the adiabatic time dependent HFB (ATDHFB) 
expression computed again in the ``cranking" approximation and given by
\cite{Gia80} 
\begin{equation}
 B_\mathrm{ATDHFB}(Q_2) = \frac{M_{-3}   (Q_2)}		    
               {M_{-1}^2 (Q_2)} \,\,,
\end{equation}
with 
\begin{equation}
\label{massparam}
M_{-n} (Q_2) = \sum_{\mu \nu} 
               \frac{ |Q^{20}_{\mu \nu}|^2 }
	            {   (E_\mu + E_\nu)^n  } \,\,.
\end{equation}
Here $Q^{20}_{\mu \nu}$ is the two-particle-zero-hole component of the quadrupole operator
$\hat{Q}_2$ in the
quasiparticle representation \cite{Ring80} and $E_\mu$ are the quasiparticle
energies obtained in the solution of the HFB equation.

In the expression for the action the collective potential $V(Q_2)$ is given 
by the HFB energy (with the 2b-KEC and REC corrections) minus the zero point 
energy (ZPE) correction $\epsilon_0 (Q_2)$ associated with the quadrupole 
motion. This ZPE correction is given by 
\begin{equation}
\epsilon_0 (Q_2) = \frac{1}{2} G(Q_2) B^{-1}_\mathrm{ATDHFB} (Q_2) \,\,,
\end{equation}
where 
\begin{equation}
 G(Q_2) = \frac{M_{-2}   (Q_2)}		    
               {2M_{-1}^2 (Q_2)} \,\,.
\end{equation}
Finally, in the expression for the action an additional parameter $E_0$ is 
introduced. This parameter can be taken as the HFB energy of the (metastable)
ground state. However, it is argued that in a quantal treatment of the problem
the ground state energy is given by the HFB energy plus the zero point energy
associated to the collective motion. To account for this fact,  the usual
recipe \cite{NTSW} is to add an estimation of the zero point energy to the HFB energy in
order to obtain $E_0$. In our calculations we have taken a zero point energy of
0.5 MeV for all the isotopes considered. 

In some SHE's around $N=176$ a weakly oblate deformed  ground state can be found.
The fission path in these nuclei obviously does not go along axial symmetric shapes through the spherical configuration. Such a nucleus rather takes triaxial shapes to reach  the prolate saddle point with the similar absolute value of the quadrupole moment. As the energies on the triaxial part of the  fission barrier are very small in comparison with the saddle point \cite{cwi05} they will contribute insignificantly to the action integral. Therefore we will neglect them in the calculation of half-lives.

To calculate the  $\alpha$-decay half-lives we use the phenomenological formula of Viola and Seaborg \cite{VS.66} 
 \begin{equation}
 \label{eq:ta}
\log T_{\alpha} [yr] = (aZ+b)(Q_{\alpha})^{-1/2} + (cZ +d) -7.5
\end{equation}
with $Z$ the atomic number of the parent nucleus.  The $Q$-factor of the
decay,   $Q_{\alpha}$,  is obtained from the calculated ground state binding energies with the experimental value $E(2,2)=-28.295674$ MeV \cite{aud03b}
\begin{equation}
Q_{\alpha} (Z,N) =  E(Z,N)-E(Z-2,N-2)- E(2,2)
\label{eq:Qalpha}
\end{equation}
The constants in Eq.~(\ref{eq:ta}) are: $a = 1.66175, b = -8.5166, c = -0.20228$ and $d = -33.9069$ 
(taken from \cite{PS.91}).
\section{Ground state properties}
\label{ssGSP}

\begin{center}
\begin{figure}
\includegraphics[angle=-90,scale=0.35]{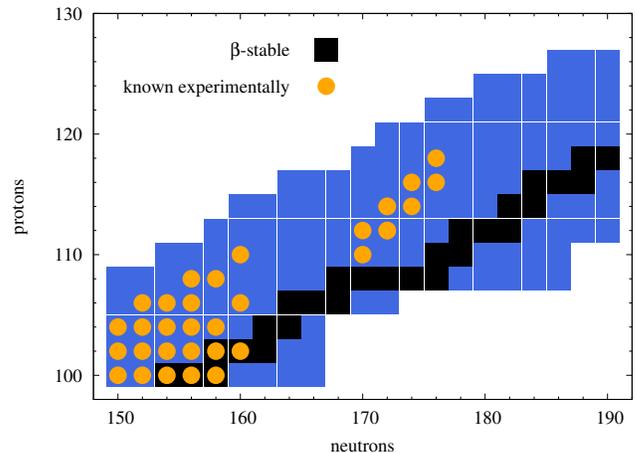}
\caption{\label{fig:all_nuclide} (Color online) Isotopes considered in this work.}
\end{figure}
\end{center}

In Fig.~\ref{fig:all_nuclide} we present an overview of all the nuclei covered in the 
present work. 
The chosen region ranges from  the heaviest trans-actinides,  well known from numerous experiments, up to  beyond the neutron magic shell  number  $N=184$ predicted by many theoretical models. The upper limit is provided by the vanishing of the two-proton separation energy. The latter approximates the  proton drip line which is determined by the one-proton separation energy with some correction due to the influence of the centrifugal barrier \cite{jac11}. The region of neutron rich isotopes beyond the $\beta$-stability line (indicated in Fig.~\ref{fig:all_nuclide} by black  squares) has been omitted as it is out of reach for the current experimental methods and can not be produced in heavy ion collisions. As it can be seen in Fig.~\ref{fig:all_nuclide} the experimentally known SHE's are located in the center of the investigated region.
 Since we consider  the ground state properties as specially relevant for the understanding of the underlying physics  
 we present these properties in the first subsection  while the following subsection is entirely devoted to 
the study of the fission barriers.

\begin{center}
\begin{figure*}
\includegraphics[angle=-90,scale=0.65]{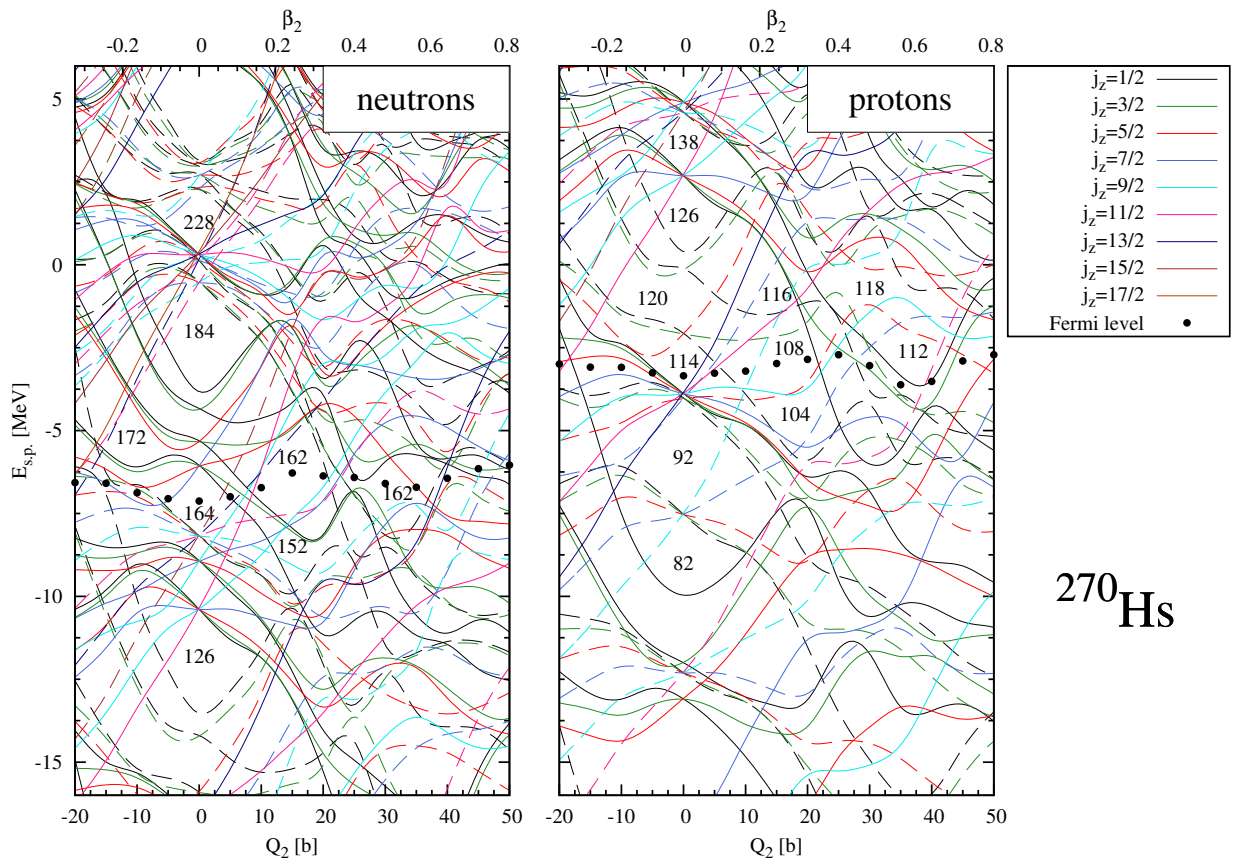}
\caption{\label{fig:spe} (Color online) Single particle energies of the nucleus $^{270}$Hs as a function of the 
deformation parameter $\beta_2$ (top x-axis) and the quadrupole moment $Q_2$ (bottom x-axis). Continuous (dashed) lines represent positive (negative) parity states
and the bullets the Fermi level.}
\end{figure*}
\end{center}

The theoretical approach discussed  in the previous section has been applied to perform a
systematic study of the properties of 160 super-heavy nuclei 
in the region $150\le N \le 190 $, $100 \le Z \le 126 $. Earlier calculations in this region
have been performed in the microscopic-macroscopic approach  \cite{mun03,kow10,gam05}, 
in the HF plus BCS with the Skyrme force in the p-h channel and a monopole pairing force 
in the p-p channel  \cite{flo75,Cwi96,Rut97,Ben98,ben99,bur04,sta05} and in the full HFB approach by \cite{ber00,ber01,ber04} 
with the Gogny force. The calculations by Berger et al. are rather similar to ours in the basic aspects. 
However, in our work the  configuration space is larger, more appropriate for fission, and we allow  
 more general HFB wave functions (with more simultaneously broken symmetries). Furthermore our study is rather detailed and systematic.

\begin{center}
\begin{figure*}
\includegraphics[angle=0,scale=1.3]{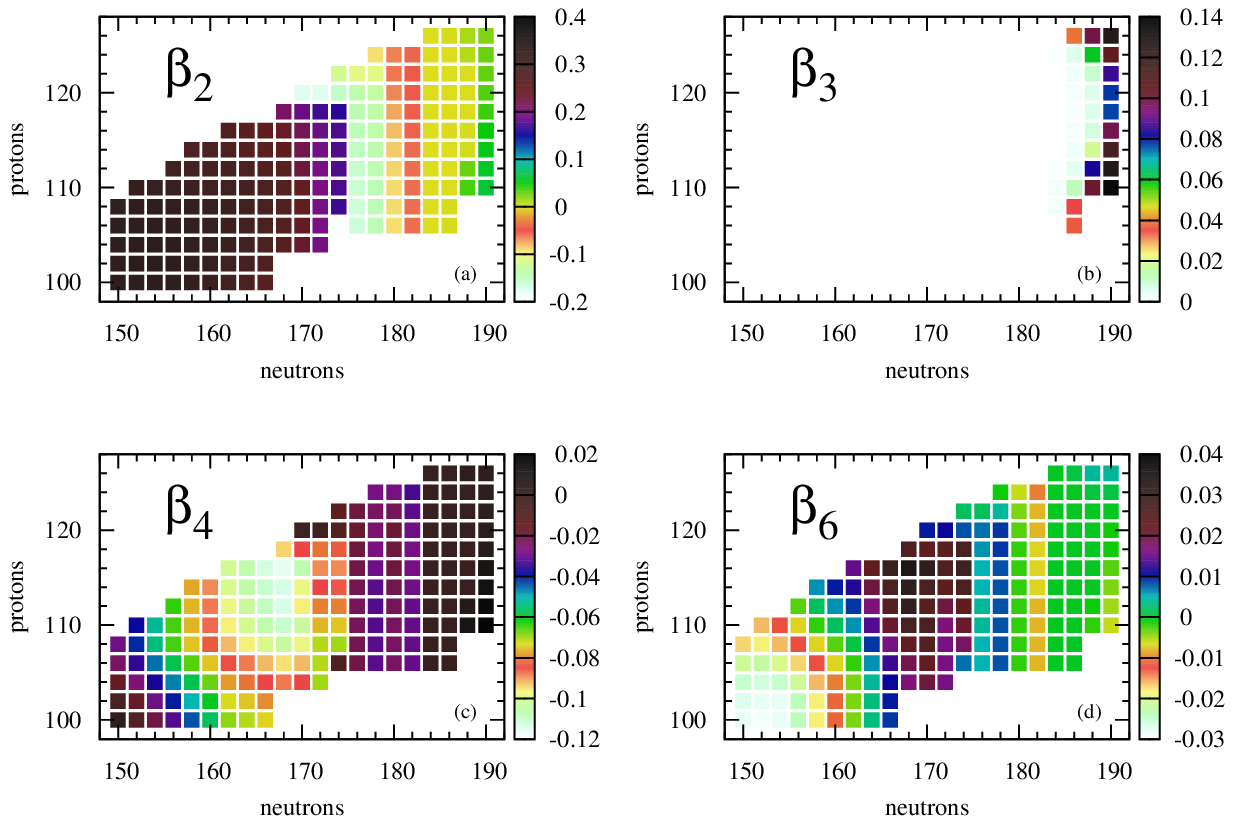}
\caption{\label{fig:beta} (Color online)  Ground state deformation parameters $\beta_2$, $\beta_3$, $\beta_4$, and $\beta_6$ for super-heavy nuclei. Notice that the deformation scale is panel dependent.}
\end{figure*}
\end{center}


 \subsection{Single particles energies}
To gain insight into the relevant physics of the SHE's we display  in Fig.~\ref{fig:spe}  the single particles energies (s.p.e.) for the nucleus $^{270}$Hs in reflection symmetric configurations as a function of the quadrupole deformation $\beta_2$ and the quadrupole moment $Q_2$. We have  chosen this nucleus as a representative of the whole region for the following reasons.  Since it is in the center of the area of the calculated  nuclei  its s.p.e. spectrum  is characteristic  for all SHE's, furthermore its proton and neutron pairing energies are small  and  the extraction of the s.p.e. from the HFB calculations is more reliable. 
The s.p.e. are obtained as usual: After solving the HFB equations the  one-body HF Hamiltonian is diagonalized in that basis. 
For the neutron system we find large spherical gaps 
at neutron numbers 164, 184 and 228; prolate gaps at 162 and oblate gaps at 172 and 178. 
In nuclei with neutron number close to 160 the spherical minimum at 164 is overwhelmed by the prolate one at 162. 
For protons there are several gaps.  The more relevant are the spherical  ones for $Z$ values 
92, 114 and 126, the spherical-oblate one at 120 and the prolate ones at 104 and 108. As we will see below many  nuclear properties can be understood just by  looking at these s.p.e.'s.


 \subsection{Deformations}
Let's now discuss some ground state properties starting with the deformation parameters
$\beta_n$. Since most of the studies on these parameters have been done in the 
macroscopic-microscopic  models we will use a definition of the $\beta_n$'s in the spirit
 of these models. The deformation parameters $\beta_n$ entering in 
$R(\theta)= R_0[1+ \sum_{n=2}^{10}\beta_n P_n(cos\theta)]$ 
are determined in such a way that the multipole moments, ($Q_\lambda, \;\lambda =2-10$), 
calculated with $R(\theta)$ and the HFB self consistent ones do coincide.  The  deformation
parameters $\beta_2$ of the calculated  SHE's are given in the third column of Table \ref{TAB1} and they are visualized in Fig.~\ref{fig:beta}a. The quadrupole 
deformation parameter $\beta_2$ barely depends on the  proton number Z and  decreases
slowly with increasing neutron number.   For all Z values the nuclei with $N<170$ have a prolate  deformation  with $\beta_2>0.25$.  In the neutron number region  $170 \le N \le 182$, the potential energy surfaces (PES) \cite{BBD.01,war06} of these nuclei present coexisting  prolate and oblate  minima. The quoted 
values for those nuclei are the ones corresponding to the deeper minimum.
 The prolate minimum is deeper in lighter nuclei with  $ N \le 174$ and $ Z \le 118$, while the  oblate one  is deeper 
  in the heavier ones. With increasing neutron number  the depth of the  prolate well becomes smaller compared to the oblate one with the oblate minimum becoming the 
lowest one around N=174 (see also Fig.~\ref{fig:Fb106-110}).  At the same time the absolute value of the quadrupole deformation parameter of the ground state decreases from $|\beta_2|=0.2$ for $N=172$  to $|\beta_2|=0.05$ for $N=182$. The barrier between the two wells diminishes and finally disappears 
at $N=184$, where the nuclei become spherical. We also observe in this region a weak Z dependency. For isotopes with   $N>184$ small prolate deformation can be observed in the ground state.
 As compared with other calculations \cite{MN.95,SM.05,BL.05,GOR.01} of SHE's our  $\beta_2$ 
 are somewhat larger but this may be connected to the slightly different methods of computation of the deformation parameters $\beta_n$.
In the presence of  coexisting  minima, one must be aware that when the restriction to axially symmetric
shapes is released one of the two minima may not be a true minimum, but rather a saddle point \cite{cwi05}. A clear example is the 
case of the heavier Ds isotopes.  In Fig.~\ref{fig:Fb106-110} one observes  two minima 
(prolate and  oblate) around $\pm 5$ b, but looking at triaxial calculations one finds that 
the prolate minimum is a saddle point.
 The gross  behavior of the deformation parameter can be qualitatively understood by looking at  Fig.~\ref{fig:spe}. Here we observe that 
shell effects favor strong prolate deformations for the nuclei with $150\leq N \leq 162 $ and 
$100\leq Z \leq 108$, i.e.,  deformation driving (down-sloping) levels are being populated and 
spherical-driving (up-sloping) depopulated. In the region $164 \leq N \leq 170 $ 
and $108\leq Z \leq 116$ neutron shell effects favor smaller prolate deformations while in 
the  proton system small  prolate as well as spherical shapes are favored. Around $N=172$ the 
oblate shapes are favored while larger  N values prefer smaller oblaticity. This effect is reinforced by the soft oblate proton gap at $Z=120$.  The spherical
shape is the privileged one by neutron shell structure at $N=184$. This coincides with the subsequent proton spherical gaps at $Z=114$, $Z=120$, and $Z=126$ which cover most of the  $N=184$ isotones considered here.

In columns 4 to 7 of Table \ref{TAB1}  as well as in Fig.~\ref{fig:beta}b,c,d we present the deformation parameters of higher multipolarities. Most of the nuclei are reflection symmetric in the ground states and only for  nuclei with $N>184$  we obtain  octupole deformed ground states independent of the proton number. M\"oller et al. \cite{MN.95} also found non-zero $\beta_3$ deformations
for the SHE's in this region.
The odd multipolarities $\beta_5, \beta_7$ and $\beta_9$, not shown 
here, are different from zero only for those nuclei whose ground state is octupole deformed, however the numerical
values are very small (e.g. $\beta_5$ values are less than 0.009).
Concerning the even deformations higher than two, see Fig.~\ref{fig:beta}c,d,  they are different from zero only for  nuclei with $\beta_2 \ne 0$, i.e., for  all nuclei with the exception of the few isotopes around  $N = 186$. 
As we can observe   there is a smaller dependence with $Z$ than with $N$.
Nuclei with $N \approx 150$ have small negative hexadecapole deformations.  For increasing $N$,   $|\beta_4|$
increases  up to $N=168$ where it reaches the largest absolute value. Large negative hexadecapole moment together with large positive quadrupole moment produces barrel-like shapes in the ground state of nuclei in this region. From here on
$|\beta_4|$ decreases rather smoothly up to $N=176$ where it sharply decreases to very small deformations
and zero values at the largest neutron number analyzed. This sharp decrease is associated to the prolate 
oblate shape transition that takes place at this neutron numbers.
 For $N$-values $150\le N \le 176$, i.e., the prolate part, there is a clear tendency with $Z$ for a 
given $N$: with increasing proton number  $|\beta_4|$ gets larger. For the oblate-spherical part, there is
almost no dependence with $Z$. 
 The behavior of the deformation parameter $\beta_6$ is quite different from $\beta_4$. For $150\le N \le 160$
we obtain negative values decreasing in absolute value as $N$ increases. Around $N=162$, $\beta_6$
becomes almost zero and for $164 \le N \le 174$ we obtain increasing values of $\beta_6$ as $N$ increases.
For $N \le 176$  we obtain rather small $\beta_6$ deformations. Our results for $\beta_4$ and $\beta_6$ are rather similar
to those of M\"oller et al \cite{MN.95} and Sobiczewski et al \cite{MH.03}.
Contrary to the $\beta_4$ and $\beta_6$ parameters, $\beta_8$ is positive in the region $150\le N \le 162$.
The nuclei with $N=164$ and 166 have either zero or very small $\beta_8$ values. The nuclei with
$168\le N \le 174$ have mostly negative $\beta_8$ values. With the exception of the $N=178$ isotopes that have
positive deformations,  all the nuclei with $N \ge 176$ have  $\beta_8$  deformations almost zero.
We have to notice that the maximal absolute values of the deformation parameters decrease with increasing multipolarity indicating a decreasing relevance.  This means that higher multipolarities can be omitted in the minimization of the energy of the ground state in non-self-consistent approximations.


\begin{longtable*}{ll ddddd ddddd dd}
\caption{Selected properties of super-heavy nuclei: ground state deformation parameters, pairing energies, $Q_{\alpha}$, 2-nucleon separation energies and half-lives for $\alpha$ emission and spontaneous fission calculated in HFB theory. The energies are given in MeV and half-lives in s.}
\label{TAB1}
\\
\hline
\\
\multicolumn{1}{c}{$Z$}&
\multicolumn{1}{c}{$N$}&
\multicolumn{1}{c}{$\beta_2$}&
\multicolumn{1}{c}{$\beta_3$}&
\multicolumn{1}{c}{$\beta_4$}&
\multicolumn{1}{c}{$\beta_6$}&
\multicolumn{1}{c}{$\beta_8$}&
\multicolumn{1}{c}{$E_N^\mathrm{PAIR}$}&
\multicolumn{1}{c}{$E_P^\mathrm{PAIR}$}&
\multicolumn{1}{c}{$Q_{\alpha}$}&
\multicolumn{1}{c}{$S_{2N}$}&
\multicolumn{1}{c}{$S_{2P}$}&
\multicolumn{1}{c}{$\log_{10}(T_{\alpha})$}&
\multicolumn{1}{c}{$\log_{10}(T_{sf})$}\\

\\
\hline
\\

 \\

100 & 150 &  0.329 &  0.000 &  0.003 & -0.028 &  0.005 &  -4.28 &  -7.31 &   7.35 &  13.20 &   8.86 &   4.04  & 13.17   \\
100 & 152 &  0.331 &  0.000 & -0.009 & -0.029 &  0.007 &  -3.85 &  -7.69 &   7.12 &  12.32 &   9.93 &   4.97  & 14.32   \\
100 & 154 &  0.332 &  0.000 & -0.021 & -0.028 &  0.010 &  -5.87 &  -8.29 &   7.02 &  11.35 &  10.87 &   5.39  & 12.82   \\
100 & 156 &  0.329 &  0.000 & -0.032 & -0.024 &  0.011 &  -6.88 &  -9.26 &   6.66 &  10.77 &  11.74 &   6.98  &  7.07   \\
100 & 158 &  0.321 &  0.000 & -0.044 & -0.018 &  0.011 &  -7.01 & -10.66 &   6.21 &  10.35 &  12.57 &   9.15  & -0.06   \\
100 & 160 &  0.310 &  0.000 & -0.056 & -0.011 &  0.009 &  -5.80 & -12.25 &   5.72 &  10.01 &  13.37 &  11.81  &  0.37   \\
100 & 162 &  0.300 &  0.000 & -0.069 & -0.004 &  0.008 &  -0.70 & -13.67 &   5.43 &   9.50 &  14.16 &  13.55  &  0.16   \\
100 & 164 &  0.285 &  0.000 & -0.071 &  0.003 &  0.005 &  -6.73 & -14.36 &   5.73 &   8.41 &  14.72 &  11.75  & -3.02   \\
100 & 166 &  0.267 &  0.000 & -0.074 &  0.009 &  0.001 &  -9.02 & -15.28 &   5.56 &   8.02 &  15.25 &  12.75  & -6.38   \\
 \\

102 & 150 &  0.331 &  0.000 & -0.006 & -0.027 &  0.005 &  -4.47 &  -6.59 &   8.12 &  14.29 &   6.98 &   1.97  &  7.63   \\
102 & 152 &  0.334 &  0.000 & -0.017 & -0.028 &  0.007 &  -3.43 &  -6.51 &   7.93 &  13.39 &   8.05 &   2.64  & 10.02   \\
102 & 154 &  0.335 &  0.000 & -0.029 & -0.028 &  0.010 &  -5.55 &  -6.77 &   7.91 &  12.34 &   9.04 &   2.71  &  8.23   \\
102 & 156 &  0.332 &  0.000 & -0.040 & -0.024 &  0.012 &  -6.62 &  -7.45 &   7.57 &  11.69 &   9.96 &   3.99  &  0.76   \\
102 & 158 &  0.325 &  0.000 & -0.050 & -0.018 &  0.011 &  -6.88 &  -8.61 &   7.09 &  11.25 &  10.86 &   5.94  &  0.57   \\
102 & 160 &  0.315 &  0.000 & -0.060 & -0.011 &  0.009 &  -5.60 &  -9.93 &   6.55 &  10.89 &  11.74 &   8.38  &  1.65   \\
102 & 162 &  0.305 &  0.000 & -0.073 & -0.004 &  0.009 &  -0.33 & -11.43 &   6.19 &  10.37 &  12.61 &  10.19  &  2.23   \\
102 & 164 &  0.293 &  0.000 & -0.075 &  0.003 &  0.005 &  -6.53 & -12.43 &   6.72 &   8.97 &  13.17 &   7.58  & -1.43   \\
102 & 166 &  0.276 &  0.000 & -0.079 &  0.010 &  0.001 &  -8.91 & -13.85 &   6.54 &   8.59 &  13.74 &   8.43  & -4.70   \\
 \\

104 & 150 &  0.331 &  0.000 & -0.016 & -0.025 &  0.005 &  -4.85 &  -4.76 &   8.91 &  15.35 &   5.10 &   0.11  & -0.41   \\
104 & 152 &  0.333 &  0.000 & -0.026 & -0.026 &  0.007 &  -3.35 &  -4.37 &   8.78 &  14.42 &   6.13 &   0.52  &  1.85   \\
104 & 154 &  0.334 &  0.000 & -0.037 & -0.025 &  0.010 &  -5.63 &  -4.53 &   8.85 &  13.32 &   7.11 &   0.30  &  1.77   \\
104 & 156 &  0.332 &  0.000 & -0.047 & -0.021 &  0.011 &  -6.78 &  -5.03 &   8.52 &  12.67 &   8.09 &   1.36  &  1.90   \\
104 & 158 &  0.325 &  0.000 & -0.057 & -0.016 &  0.010 &  -6.92 &  -5.99 &   7.98 &  12.23 &   9.07 &   3.24  &  2.03   \\
104 & 160 &  0.317 &  0.000 & -0.067 & -0.010 &  0.009 &  -5.39 &  -7.17 &   7.40 &  11.83 &  10.01 &   5.47  &  3.86   \\
104 & 162 &  0.309 &  0.000 & -0.078 & -0.004 &  0.009 &  -0.06 &  -8.76 &   6.99 &  11.30 &  10.94 &   7.22  &  5.36   \\
104 & 164 &  0.299 &  0.000 & -0.081 &  0.004 &  0.005 &  -6.28 &  -9.99 &   7.75 &   9.61 &  11.58 &   4.09  &  1.11   \\
104 & 166 &  0.285 &  0.000 & -0.085 &  0.011 &  0.001 &  -8.87 & -11.54 &   7.54 &   9.18 &  12.17 &   4.91  & -2.64   \\
 \\

106 & 150 &  0.329 &  0.000 & -0.024 & -0.021 &  0.004 &  -5.50 &  -6.23 &  10.04 &  16.27 &   2.91 &  -2.43  & -3.01   \\
106 & 152 &  0.331 &  0.000 & -0.035 & -0.022 &  0.006 &  -3.81 &  -5.49 &   9.98 &  15.41 &   3.90 &  -2.28  & -0.11   \\
106 & 154 &  0.331 &  0.000 & -0.046 & -0.020 &  0.009 &  -6.10 &  -5.02 &  10.03 &  14.37 &   4.95 &  -2.41  &  0.70   \\
106 & 156 &  0.328 &  0.000 & -0.057 & -0.017 &  0.010 &  -7.21 &  -5.03 &   9.61 &  13.74 &   6.02 &  -1.26  &  1.20   \\
106 & 158 &  0.324 &  0.000 & -0.066 & -0.012 &  0.010 &  -7.08 &  -5.22 &   8.97 &  13.31 &   7.10 &   0.63  &  2.68   \\
106 & 160 &  0.317 &  0.000 & -0.075 & -0.007 &  0.009 &  -5.25 &  -5.66 &   8.31 &  12.89 &   8.16 &   2.82  &  5.75   \\
106 & 162 &  0.310 &  0.000 & -0.084 & -0.002 &  0.009 &  -0.01 &  -6.39 &   7.81 &  12.33 &   9.19 &   4.65  &  9.30   \\
106 & 164 &  0.301 &  0.000 & -0.088 &  0.006 &  0.004 &  -5.87 &  -7.37 &   8.77 &  10.34 &   9.92 &   1.27  &  3.72   \\
106 & 166 &  0.288 &  0.000 & -0.092 &  0.013 &  0.001 &  -8.28 &  -8.84 &   8.46 &   9.92 &  10.66 &   2.30  & -0.11   \\
106 & 168 &  0.272 &  0.000 & -0.095 &  0.019 & -0.002 &  -8.81 & -10.72 &   8.03 &   9.61 &  11.24 &   3.82  & -4.13   \\
106 & 170 &  0.246 &  0.000 & -0.092 &  0.022 & -0.004 &  -8.29 & -13.34 &   7.69 &   9.37 &  11.82 &   5.12  & -5.79   \\
106 & 172 &  0.196 &  0.000 & -0.071 &  0.016 & -0.002 &  -6.31 & -17.22 &   7.18 &   9.30 &  12.40 &   7.23  & -5.47   \\
 \\

108 & 150 &  0.324 &  0.000 & -0.029 & -0.017 &  0.004 &  -6.11 &  -7.61 &  11.11 &  17.08 &   0.92 &  -4.45  & -5.43   \\
108 & 152 &  0.327 &  0.000 & -0.041 & -0.018 &  0.006 &  -4.46 &  -6.01 &  11.06 &  16.32 &   1.83 &  -4.34  & -2.77   \\
108 & 154 &  0.327 &  0.000 & -0.054 & -0.016 &  0.008 &  -6.76 &  -4.49 &  11.10 &  15.37 &   2.83 &  -4.43  & -0.95   \\
108 & 156 &  0.325 &  0.000 & -0.065 & -0.012 &  0.009 &  -7.82 &  -3.47 &  10.66 &  14.81 &   3.90 &  -3.38  &  0.35   \\
108 & 158 &  0.321 &  0.000 & -0.075 & -0.007 &  0.009 &  -7.40 &  -3.05 &  10.01 &  14.39 &   4.98 &  -1.71  &  2.64   \\
108 & 160 &  0.316 &  0.000 & -0.083 & -0.003 &  0.009 &  -5.33 &  -2.89 &   9.36 &  13.96 &   6.05 &   0.14  &  6.49   \\
108 & 162 &  0.310 &  0.000 & -0.092 &  0.001 &  0.008 &  -0.02 &  -3.04 &   8.84 &  13.41 &   7.13 &   1.76  & 12.83   \\
108 & 164 &  0.300 &  0.000 & -0.096 &  0.009 &  0.004 &  -5.29 &  -4.26 &  10.02 &  11.15 &   7.94 &  -1.74  &  5.89   \\
108 & 166 &  0.290 &  0.000 & -0.100 &  0.016 &  0.000 &  -7.58 &  -5.88 &   9.65 &  10.71 &   8.73 &  -0.71  &  2.66   \\
108 & 168 &  0.276 &  0.000 & -0.103 &  0.022 & -0.003 &  -8.16 &  -8.10 &   9.22 &  10.35 &   9.47 &   0.56  & -1.41   \\
108 & 170 &  0.250 &  0.000 & -0.098 &  0.025 & -0.004 &  -7.84 & -11.26 &   8.68 &  10.15 &  10.25 &   2.29  & -3.84   \\
108 & 172 &  0.195 &  0.000 & -0.071 &  0.016 & -0.002 &  -6.07 & -15.49 &   8.08 &   9.97 &  10.92 &   4.40  & -3.54   \\
108 & 174 &  0.170 &  0.000 & -0.070 &  0.018 & -0.003 &  -3.95 & -17.17 &   7.32 &  10.06 &  11.56 &   7.45  & -4.43   \\
108 & 176 & -0.162 &  0.000 & -0.022 &  0.005 &  0.002 &  -4.53 & -19.62 &   6.72 &  10.02 &  11.99 &  10.21  &  6.25   \\
108 & 178 & -0.146 &  0.000 & -0.031 &  0.008 &  0.006 &  -0.01 & -20.36 &   6.49 &   9.82 &  12.67 &  11.37  & 14.21   \\
108 & 180 & -0.090 &  0.000 & -0.024 & -0.002 &  0.001 &  -3.12 & -22.30 &   6.69 &   8.94 &  13.23 &  10.36  &  5.76   \\
108 & 182 & -0.060 &  0.000 & -0.030 & -0.008 & -0.001 &   0.00 & -23.23 &   6.44 &   8.63 &  13.83 &  11.63  & 14.10   \\
108 & 184 & -0.037 &  0.000 &  0.000 &  0.000 &  0.000 &   0.00 & -23.95 &   7.06 &   7.41 &  14.28 &   8.61  &  8.40   \\
108 & 186 &  0.034 & -0.062 &  0.005 &  0.000 &  0.000 &  -5.76 & -23.57 &   7.00 &   7.02 &  14.50 &   8.88  &  0.68   \\
 \\

110 & 154 &  0.319 &  0.000 & -0.052 & -0.012 &  0.006 &  -7.43 &  -8.77 &  12.31 &  16.15 &   0.62 &  -6.48  & -4.27   \\
110 & 156 &  0.314 &  0.000 & -0.065 & -0.007 &  0.007 &  -8.68 &  -7.91 &  11.97 &  15.71 &   1.52 &  -5.78  & -2.38   \\
110 & 158 &  0.310 &  0.000 & -0.076 & -0.002 &  0.007 &  -8.12 &  -7.25 &  11.41 &  15.37 &   2.50 &  -4.55  &  0.14   \\
110 & 160 &  0.306 &  0.000 & -0.087 &  0.002 &  0.006 &  -5.66 &  -6.75 &  10.79 &  15.01 &   3.55 &  -3.09  &  3.94   \\
110 & 162 &  0.303 &  0.000 & -0.096 &  0.006 &  0.007 &  -0.03 &  -6.39 &  10.32 &  14.43 &   4.57 &  -1.90  &  9.02   \\
110 & 164 &  0.293 &  0.000 & -0.101 &  0.014 &  0.003 &  -4.93 &  -6.71 &  11.41 &  12.32 &   5.74 &  -4.55  &  5.87   \\
110 & 166 &  0.283 &  0.000 & -0.106 &  0.021 & -0.001 &  -6.63 &  -7.26 &  10.77 &  11.79 &   6.82 &  -3.04  &  3.63   \\
110 & 168 &  0.271 &  0.000 & -0.110 &  0.027 & -0.004 &  -6.90 &  -8.28 &  10.14 &  11.34 &   7.81 &  -1.42  &  1.94   \\
110 & 170 &  0.247 &  0.000 & -0.103 &  0.027 & -0.005 &  -7.25 & -10.38 &   9.63 &  10.86 &   8.52 &   0.01  & -1.98   \\
110 & 172 &  0.197 &  0.000 & -0.076 &  0.017 & -0.002 &  -5.86 & -13.21 &   8.99 &  10.79 &   9.34 &   1.98  & -0.76   \\
110 & 174 &  0.175 &  0.000 & -0.075 &  0.021 & -0.004 &  -3.28 & -14.84 &   8.24 &  10.72 &  10.00 &   4.57  & -1.87   \\
110 & 176 & -0.160 &  0.000 & -0.023 &  0.004 &  0.002 &  -4.45 & -18.70 &   7.72 &  10.58 &  10.56 &   6.58  & -0.48   \\
110 & 178 & -0.145 &  0.000 & -0.032 &  0.008 &  0.006 &   0.00 & -19.31 &   7.24 &  10.50 &  11.24 &   8.63  &  4.61   \\
110 & 180 & -0.090 &  0.000 & -0.024 & -0.002 &  0.001 &  -3.08 & -21.19 &   7.53 &   9.53 &  11.83 &   7.37  &  5.67   \\
110 & 182 & -0.060 &  0.000 & -0.030 & -0.008 & -0.001 &  -0.01 & -22.18 &   7.24 &   9.23 &  12.43 &   8.63  & 13.90   \\
110 & 184 &  0.000 & -0.002 &  0.000 &  0.000 &  0.000 &   0.00 & -23.43 &   8.03 &   7.84 &  12.86 &   5.36  &  6.25   \\
110 & 186 &  0.035 & -0.035 &  0.004 &  0.000 &  0.000 &  -6.59 & -23.04 &   8.29 &   7.15 &  12.99 &   4.39  &-10.25   \\
 \\

112 & 158 &  0.297 &  0.000 & -0.076 &  0.002 &  0.004 &  -8.73 &  -9.36 &  12.07 &  16.32 &   0.86 &  -5.43  & -2.42   \\
112 & 160 &  0.295 &  0.000 & -0.088 &  0.006 &  0.005 &  -6.01 &  -8.62 &  11.50 &  15.94 &   1.79 &  -4.18  &  1.58   \\
112 & 162 &  0.294 &  0.000 & -0.098 &  0.009 &  0.005 &  -0.04 &  -7.98 &  11.17 &  15.34 &   2.70 &  -3.41  &  6.17   \\
112 & 164 &  0.283 &  0.000 & -0.103 &  0.017 &  0.001 &  -4.70 &  -7.77 &  12.10 &  13.50 &   3.88 &  -5.50  &  4.78   \\
112 & 166 &  0.273 &  0.000 & -0.109 &  0.025 & -0.002 &  -6.09 &  -7.78 &  11.52 &  12.90 &   4.99 &  -4.23  &  3.56   \\
112 & 168 &  0.262 &  0.000 & -0.113 &  0.031 & -0.005 &  -6.12 &  -8.13 &  10.94 &  12.37 &   6.02 &  -2.86  &  2.87   \\
112 & 170 &  0.253 &  0.000 & -0.114 &  0.035 & -0.007 &  -6.00 &  -8.62 &  10.32 &  11.96 &   7.12 &  -1.27  & -0.27   \\
112 & 172 &  0.211 &  0.000 & -0.087 &  0.024 & -0.004 &  -5.83 & -10.26 &   9.46 &  11.72 &   8.05 &   1.19  &  2.36   \\
112 & 174 &  0.191 &  0.000 & -0.086 &  0.026 & -0.005 &  -1.90 & -11.45 &   8.76 &  11.49 &   8.82 &   3.46  &  3.11   \\
112 & 176 & -0.157 &  0.000 & -0.023 &  0.004 &  0.002 &  -4.34 & -17.21 &   8.60 &  10.88 &   9.12 &   4.01  &  8.14   \\
112 & 178 & -0.144 &  0.000 & -0.032 &  0.008 &  0.006 &  -0.01 & -17.75 &   8.00 &  11.18 &   9.80 &   6.25  & 16.60   \\
112 & 180 & -0.088 &  0.000 & -0.023 & -0.002 &  0.001 &  -3.07 & -19.53 &   8.36 &  10.14 &  10.41 &   4.88  &  7.05   \\
112 & 182 & -0.060 &  0.000 & -0.030 & -0.008 & -0.001 &   0.00 & -20.57 &   8.02 &   9.87 &  11.05 &   6.17  & 15.09   \\
112 & 184 &  0.000 & -0.001 &  0.000 &  0.000 &  0.000 &   0.00 & -21.93 &   8.82 &   8.43 &  11.64 &   3.25  &  7.29   \\
112 & 186 &  0.052 & -0.007 &  0.005 &  0.001 &  0.001 &  -7.35 & -20.94 &   9.80 &   6.86 &  11.35 &   0.18  & -8.67   \\
112 & 188 &  0.015 & -0.072 &  0.004 & -0.001 &  0.000 & -11.05 & -22.03 &   8.68 &   8.27 &  12.86 &   3.73  & -8.10   \\
112 & 190 &  0.028 & -0.123 &  0.011 & -0.005 &  0.000 & -10.70 & -21.64 &   9.01 &   6.43 &  13.26 &   2.62  &-14.23   \\
 \\

114 & 160 &  0.284 &  0.000 & -0.089 &  0.010 &  0.003 &  -6.36 &  -8.58 &  12.08 &  16.84 &   0.28 &  -4.90  & -0.80   \\
114 & 162 &  0.285 &  0.000 & -0.100 &  0.012 &  0.004 &  -0.10 &  -7.72 &  11.87 &  16.15 &   1.09 &  -4.44  &  3.72   \\
114 & 164 &  0.272 &  0.000 & -0.105 &  0.020 &  0.000 &  -4.62 &  -6.79 &  12.53 &  14.68 &   2.27 &  -5.85  &  3.62   \\
114 & 166 &  0.262 &  0.000 & -0.110 &  0.027 & -0.003 &  -5.88 &  -6.35 &  12.05 &  13.98 &   3.35 &  -4.84  &  3.27   \\
114 & 168 &  0.251 &  0.000 & -0.113 &  0.032 & -0.005 &  -6.00 &  -6.29 &  11.56 &  13.39 &   4.37 &  -3.74  &  3.65   \\
114 & 170 &  0.231 &  0.000 & -0.106 &  0.031 & -0.006 &  -6.70 &  -6.67 &  10.97 &  12.96 &   5.37 &  -2.33  &  1.60   \\
114 & 172 &  0.210 &  0.000 & -0.094 &  0.028 & -0.005 &  -5.48 &  -7.17 &  10.28 &  12.65 &   6.30 &  -0.53  &  5.00   \\
114 & 174 &  0.190 &  0.000 & -0.091 &  0.030 & -0.006 &  -1.13 &  -8.09 &   9.69 &  12.31 &   7.12 &   1.17  &  6.38   \\
114 & 176 & -0.153 &  0.000 & -0.022 &  0.005 &  0.002 &  -4.32 & -15.07 &   9.74 &  11.44 &   7.68 &   1.02  & 10.60   \\
114 & 178 & -0.141 &  0.000 & -0.031 &  0.008 &  0.006 &  -0.01 & -15.58 &   8.76 &  11.86 &   8.36 &   4.18  & 18.54   \\
114 & 180 & -0.085 &  0.000 & -0.022 & -0.002 &  0.000 &  -3.10 & -17.31 &   9.13 &  10.81 &   9.03 &   2.92  &  9.45   \\
114 & 182 & -0.059 &  0.000 & -0.029 & -0.007 & -0.001 &  -0.01 & -18.41 &   8.78 &  10.49 &   9.65 &   4.11  & 17.10   \\
114 & 184 &  0.000 &  0.000 &  0.000 &  0.000 &  0.000 &   0.00 & -19.84 &   9.68 &   8.97 &  10.19 &   1.20  &  8.59   \\
114 & 186 &  0.052 & -0.005 &  0.004 &  0.001 &  0.001 &  -7.66 & -18.85 &  10.60 &   7.51 &  10.84 &  -1.39  & -6.76   \\
114 & 188 &  0.032 & -0.064 &  0.006 &  0.000 &  0.000 & -10.96 & -20.07 &   8.81 &   8.65 &  11.22 &   4.00  & -7.58   \\
114 & 190 & -0.019 & -0.072 &  0.003 & -0.001 &  0.000 & -16.36 & -19.89 &   9.91 &   7.17 &  11.96 &   0.52  &-12.52   \\
 \\

116 & 164 &  0.262 &  0.000 & -0.106 &  0.023 &  0.000 &  -4.71 &  -3.91 &  13.12 &  15.74 &   0.50 &  -6.50  &  1.17   \\
116 & 166 &  0.252 &  0.000 & -0.112 &  0.030 & -0.004 &  -5.81 &  -2.26 &  12.76 &  15.04 &   1.56 &  -5.78  &  3.84   \\
116 & 168 &  0.242 &  0.000 & -0.115 &  0.035 & -0.006 &  -5.83 &  -1.49 &  12.32 &  14.42 &   2.59 &  -4.87  &  5.14   \\
116 & 170 &  0.231 &  0.000 & -0.114 &  0.037 & -0.007 &  -6.02 &  -1.32 &  11.74 &  13.97 &   3.60 &  -3.59  &  2.78   \\
116 & 172 &  0.189 &  0.000 & -0.089 &  0.027 & -0.005 &  -5.08 &  -4.87 &  11.20 &  13.50 &   4.45 &  -2.31  &  6.02   \\
116 & 174 &  0.188 &  0.000 & -0.095 &  0.033 & -0.007 &  -0.44 &  -4.55 &  10.67 &  13.18 &   5.32 &  -0.96  &  8.01   \\
116 & 176 & -0.147 &  0.000 & -0.021 &  0.005 &  0.002 &  -4.30 & -12.14 &  10.66 &  12.32 &   6.20 &  -0.93  & 13.72   \\
116 & 178 & -0.138 &  0.000 & -0.029 &  0.008 &  0.006 &  -0.01 & -12.67 &   9.55 &  12.55 &   6.89 &   2.26  & 21.11   \\
116 & 180 & -0.081 &  0.000 & -0.020 & -0.002 &  0.000 &  -3.17 & -14.50 &   9.90 &  11.51 &   7.59 &   1.20  & 12.61   \\
116 & 182 & -0.057 &  0.000 & -0.028 & -0.007 & -0.001 &  -0.01 & -15.68 &   9.57 &  11.14 &   8.24 &   2.20  & 19.87   \\
116 & 184 &  0.000 &  0.000 &  0.000 &  0.000 &  0.000 &   0.00 & -17.19 &  10.51 &   9.55 &   8.82 &  -0.53  & 11.02   \\
116 & 186 &  0.051 & -0.008 &  0.003 &  0.001 &  0.001 &  -7.91 & -16.20 &  11.31 &   8.17 &   9.48 &  -2.58  & -4.28   \\
116 & 188 &  0.033 & -0.039 &  0.004 &  0.000 &  0.000 & -12.40 & -17.04 &   9.89 &   8.93 &   9.76 &   1.23  & -6.10   \\
116 & 190 &  0.015 & -0.078 &  0.004 & -0.002 &  0.000 & -15.32 & -18.02 &  10.38 &   8.16 &  10.75 &  -0.17  & -9.85   \\
 \\

118 & 170 &  0.208 &  0.000 & -0.095 &  0.029 & -0.005 &  -7.20 &  -2.98 &  13.37 &  15.09 &   0.96 &  -6.47  &  1.39   \\
118 & 172 &  0.171 &  0.000 & -0.081 &  0.024 & -0.004 &  -4.96 &  -3.14 &  12.57 &  14.77 &   2.23 &  -4.86  &  6.27   \\
118 & 174 &  0.168 &  0.000 & -0.088 &  0.030 & -0.006 &  -1.39 &  -2.40 &  11.92 &  14.15 &   3.20 &  -3.44  &  9.44   \\
118 & 176 & -0.143 &  0.000 & -0.020 &  0.006 &  0.002 &  -4.32 &  -8.21 &  11.33 &  13.77 &   4.65 &  -2.04  & 17.87   \\
118 & 178 & -0.135 &  0.000 & -0.027 &  0.009 &  0.005 &  -0.01 &  -8.89 &  10.42 &  13.23 &   5.33 &   0.34  & 24.12   \\
118 & 180 & -0.075 &  0.000 & -0.020 & -0.002 &  0.000 &  -3.30 & -11.26 &  10.70 &  12.27 &   6.09 &  -0.42  & 16.37   \\
118 & 182 & -0.056 &  0.000 & -0.028 & -0.007 & -0.001 &  -0.01 & -12.42 &  10.40 &  11.81 &   6.76 &   0.40  & 22.89   \\
118 & 184 &  0.000 &  0.000 &  0.000 &  0.000 &  0.000 &   0.00 & -14.04 &  11.37 &  10.17 &   7.38 &  -2.14  & 13.92   \\
118 & 186 &  0.051 & -0.008 &  0.001 &  0.001 &  0.000 &  -8.12 & -13.06 &  12.15 &   8.77 &   7.98 &  -3.95  & -2.03   \\
118 & 188 &  0.049 & -0.054 &  0.006 &  0.001 &  0.000 & -11.50 & -13.75 &  10.73 &   9.59 &   8.64 &  -0.50  & -4.10   \\
118 & 190 &  0.016 & -0.054 &  0.003 & -0.001 &  0.000 & -16.98 & -14.30 &  10.80 &   8.86 &   9.34 &  -0.69  & -7.35   \\
 \\

120 & 172 & -0.176 &  0.000 & -0.002 &  0.011 &  0.000 &  -3.70 &  -2.47 &  12.63 &  15.84 &   0.90 &  -4.46  & 12.15   \\
120 & 174 & -0.154 &  0.000 & -0.010 &  0.008 &  0.001 &  -5.41 &  -1.89 &  12.26 &  15.14 &   1.89 &  -3.65  & 15.82   \\
120 & 176 & -0.138 &  0.000 & -0.018 &  0.007 &  0.003 &  -4.35 &  -3.44 &  11.73 &  14.68 &   2.80 &  -2.43  & 19.57   \\
120 & 178 & -0.130 &  0.000 & -0.025 &  0.009 &  0.005 &  -0.01 &  -4.81 &  11.53 &  13.97 &   3.54 &  -1.95  & 24.62   \\
120 & 180 & -0.066 &  0.000 & -0.019 & -0.003 &  0.000 &  -3.56 &  -7.97 &  11.60 &  13.16 &   4.43 &  -2.12  & 19.80   \\
120 & 182 & -0.055 &  0.000 & -0.029 & -0.008 & -0.001 &  -0.01 &  -8.83 &  11.29 &  12.58 &   5.20 &  -1.36  & 25.60   \\
120 & 184 &  0.000 &  0.000 &  0.000 &  0.000 &  0.000 &   0.00 & -10.57 &  12.30 &  10.80 &   5.83 &  -3.74  & 16.29   \\
120 & 186 &  0.050 & -0.012 &  0.001 &  0.001 &  0.000 &  -8.16 &  -9.69 &  13.05 &   9.42 &   6.48 &  -5.33  &  0.27   \\
120 & 188 &  0.048 & -0.059 &  0.005 &  0.001 &  0.000 & -11.37 & -10.34 &  11.57 &  10.25 &   7.14 &  -2.05  & -1.88   \\
120 & 190 &  0.016 & -0.052 &  0.002 &  0.000 &  0.000 & -17.16 & -10.62 &  11.61 &   9.55 &   7.83 &  -2.15  & -5.01   \\
 \\

122 & 176 & -0.109 &  0.000 & -0.020 &  0.002 &  0.002 &  -5.35 &  -4.68 &  13.02 &  15.62 &   0.60 &  -4.75  & 17.54   \\
122 & 178 & -0.109 &  0.000 & -0.024 &  0.004 &  0.003 &  -1.11 &  -4.56 &  12.93 &  14.77 &   1.40 &  -4.56  & 19.76   \\
122 & 180 & -0.062 &  0.000 & -0.023 & -0.004 &  0.000 &  -3.62 &  -4.98 &  12.50 &  14.40 &   2.64 &  -3.64  & 21.94   \\
122 & 182 & -0.055 &  0.000 & -0.031 & -0.008 & -0.001 &   0.00 &  -5.24 &  12.27 &  13.39 &   3.45 &  -3.13  & 28.05   \\
122 & 184 &  0.000 &  0.000 &  0.000 &  0.000 &  0.000 &   0.00 &  -7.19 &  13.33 &  11.52 &   4.17 &  -5.38  & 18.02   \\
122 & 186 &  0.049 & -0.036 &  0.000 &  0.001 &  0.000 &  -7.63 &  -6.64 &  13.52 &  10.61 &   5.36 &  -5.76  &  3.69   \\
122 & 188 &  0.031 & -0.059 &  0.002 &  0.001 &  0.000 & -11.75 &  -7.17 &  12.10 &  10.84 &   5.95 &  -2.74  &  1.10   \\
122 & 190 &  0.015 & -0.068 &  0.002 &  0.000 &  0.000 & -16.17 &  -7.55 &  12.47 &   9.88 &   6.28 &  -3.58  & -3.44   \\
 \\

124 & 180 & -0.065 &  0.000 & -0.028 & -0.006 &  0.000 &  -3.44 &  -1.51 &  13.59 &  15.21 &   0.31 &  -5.40  & 20.40   \\
124 & 182 & -0.058 &  0.000 & -0.033 & -0.010 & -0.001 &   0.00 &  -1.15 &  13.76 &  14.23 &   1.15 &  -5.73  & 29.94   \\
124 & 184 &  0.000 & -0.001 &  0.000 &  0.000 &  0.000 &   0.00 &  -4.30 &  14.62 &  12.53 &   2.16 &  -7.32  & 17.90   \\
124 & 186 &  0.032 & -0.050 &  0.001 &  0.002 &  0.000 &  -6.81 &  -4.27 &  13.95 &  12.19 &   3.74 &  -6.09  &  9.63   \\
124 & 188 &  0.015 & -0.068 &  0.002 &  0.002 &  0.000 & -11.72 &  -4.48 &  13.55 &  11.01 &   3.91 &  -5.32  &  2.93   \\
124 & 190 & -0.002 & -0.078 &  0.002 &  0.001 &  0.000 & -16.02 &  -4.73 &  14.21 &  10.18 &   4.21 &  -6.58  & -2.94   \\
 \\

126 & 186 &  0.015 & -0.073 &  0.001 &  0.003 &  0.000 &  -6.38 &  -1.22 &  15.04 &  13.53 &   1.07 &  -7.59  &  9.82   \\
126 & 188 & -0.002 & -0.086 &  0.002 &  0.004 &  0.000 & -11.40 &  -1.03 &  16.13 &  11.10 &   1.16 &  -9.37  &  1.03   \\
126 & 190 &  0.011 & -0.121 &  0.003 &  0.004 &  0.000 & -13.12 &  -0.48 &  16.05 &  11.09 &   2.07 &  -9.25  & -4.06   \\
 \\

\end{longtable*}

 \subsection{Pairing energies}
 
 \begin{center}
\begin{figure}
\includegraphics[angle=0,scale=1.3]{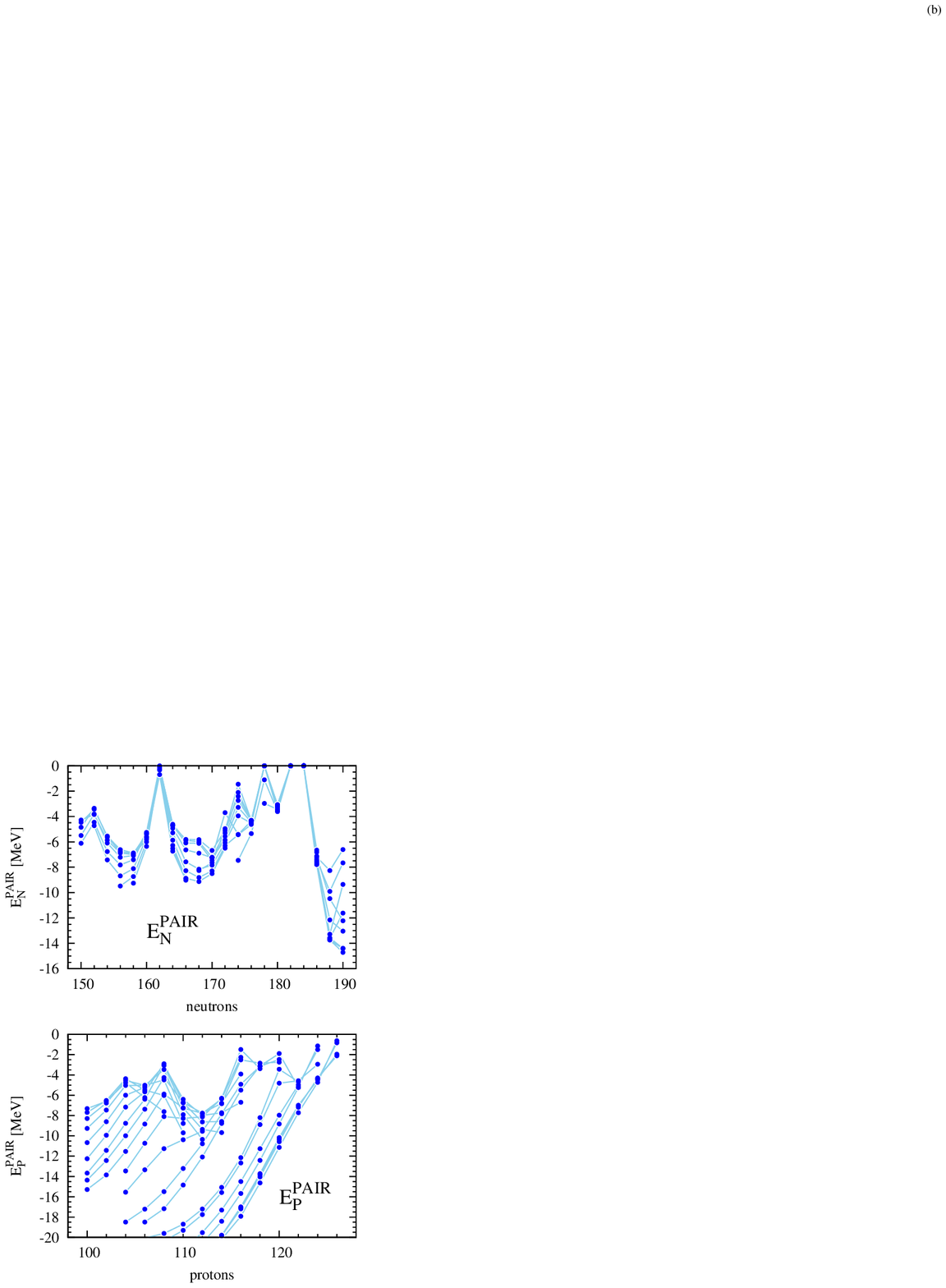}
\caption{\label{fig:pair} (Color online) Ground state proton and neutron  pairing energies for super-heavy nuclei.}
\end{figure}
\end{center}
In the HFB approach \cite{Ang01}, the pairing energy is given by
\begin{equation}
E^\mathrm{PAIR}=  - \frac{1}{2}  {\displaystyle Tr}  \left (\Delta {\kappa}^* \right),
\end{equation}
with
\begin{equation}
 \Delta_{k_1k_2} = \frac{1}{2} \sum_{k_3k_4} \bar{v}_{k_1k_2k_3k_4}\kappa_{k_3k_4},
\end{equation}
  the pairing field and 
  \begin{equation}
   \kappa_{k_1,k_2} = \langle HFB| c_{k_2}c_{k_1}|HFB \rangle,
\end{equation}   
 the pairing tensor. 

The neutron pairing energies of each nucleus are given in the eighth column of Table \ref{TAB1}  and in the upper panel of Fig.~\ref{fig:pair}.  
The general behavior of these results can be easily understood  by looking again at Fig.~\ref{fig:spe}. 
We know that small pairing energies correspond to situations of low level density of the neutron 
s.p.e. In particular we expect zero neutron pairing  energies at $N=162, 178, 182$ and 184. 
Notice that  the fact that for $N=162$ the pairing energies are zero is consistent with the prolate 
deformation of these nuclei and for $N=178$ with the oblate one.  For $N=182$ and 184 we have  spherical nuclei. Appropriate
 neutron shell gaps can be found in Fig.~\ref{fig:spe} for the mentioned cases. For the same reason large pairing energies are associated with high
level density, for example, we find large pairing energies for  $N=158, 168$ and 190. We observe that the 
general pattern,  qualitatively, does not depend  on the proton number. The proton pairing energies 
are given in the ninth column of Table \ref{TAB1}  and the lower  panel of Fig.~\ref{fig:pair}.
In this case the general pattern looks more complicated. For the same proton number some nuclei are prolate, some oblate and some  spherical and therefore only part of them fit to the proton shell gaps in Fig.~\ref{fig:spe}. 
Again the small pairing energies found at $Z$ values 104, 108, 116, 120 and 126 
have to do  with the low proton level density found in Fig.~\ref{fig:spe}  for those proton number. 
Concerning the behavior for the different isotopes, for example for $Z=112$, from $N=158$ up to 
$N=174$ we have proton pairing energies around $-9$ MeV and from $N=176$ on around $-20$ MeV.  The reason 
for this change is that in the first interval  the nuclei are prolate and in the second either oblate or spherical.  
Similar arguments apply for other cases. It is interesting to notice that with the exception of 
the  few nuclei with $ N \ge 188$ none  of the nuclei studied has absolute values of the neutron 
pairing energies larger than 9 MeV. On the other hand,  many of the analyzed nuclei have  proton 
pairing energies  much larger than 10 MeV.


 \subsection{$Q_{\alpha}$ values}
 
\begin{center}
\begin{figure}
\includegraphics[angle=0,scale=1.15]{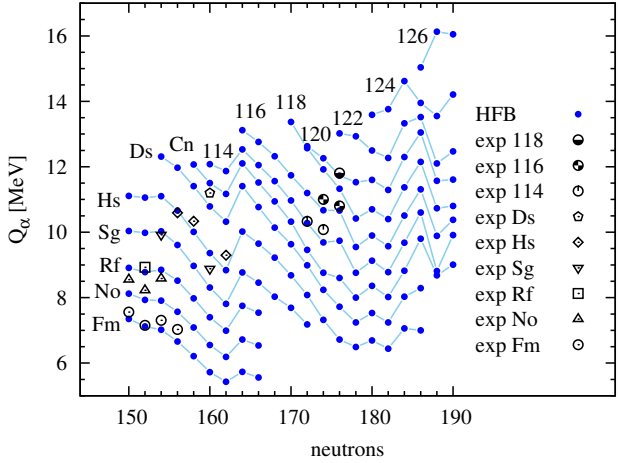}
\caption{\label{fig:Qalpha}  (Color online) $Q_\alpha$ values, theoretical values  marked by dots are compared with  experimental data \cite{aud03b,oga07}.
The different lines correspond to the indicated isotope}
\label{fig:qalpha}
\end{figure}
\end{center}
  In Fig.~\ref{fig:Qalpha} the $Q_{\alpha}$ values  are plotted as a function of the neutron number
[see Eq. (\ref{eq:Qalpha})]. The exact numerical values are also given in column 10 of Table \ref{TAB1}.
  The patterns displayed by the different isotopes are easily  understood just looking at  Eq. (\ref{eq:Qalpha}): minima  appear when the mother nucleus are more bound than the average and maxima correspond to more bound daughter nuclei. Thus
  we observe  minima corresponding to the
neutron numbers for which energy gaps appear in the single particle energies of Fig.~\ref{fig:spe}, namely 
162, 178 and 182-184. For  the $N $ values  $164$ and $172$ there is no structure because the energy gaps that
 one finds for these neutron numbers are not large enough to provide energy minima at those deformations, see column
three of Table \ref{TAB1}. Experimental values \cite{aud03b,oga07} for some isotopes of the nuclei Fm, No, Rf, Sg, Hs, Ds, 114 116 and 120
are also displayed in the Figure.  The agreement between theory and experiment is very satisfactory,
in most of the cases we obtain a quantitative agreement and for the others at least the tendency is the right one.
Concerning the proton dependence we observe a similar situation, we find large energy spacing for the lines
whose Z numbers correspond to energy gaps in the single particle diagram of Fig.~\ref{fig:spe}, for example, 
Hs, 116, 120, etc.


 \subsection{Two-nucleon separation energies}
 
\begin{center}
\begin{figure}
\includegraphics[angle=0,scale=1.3]{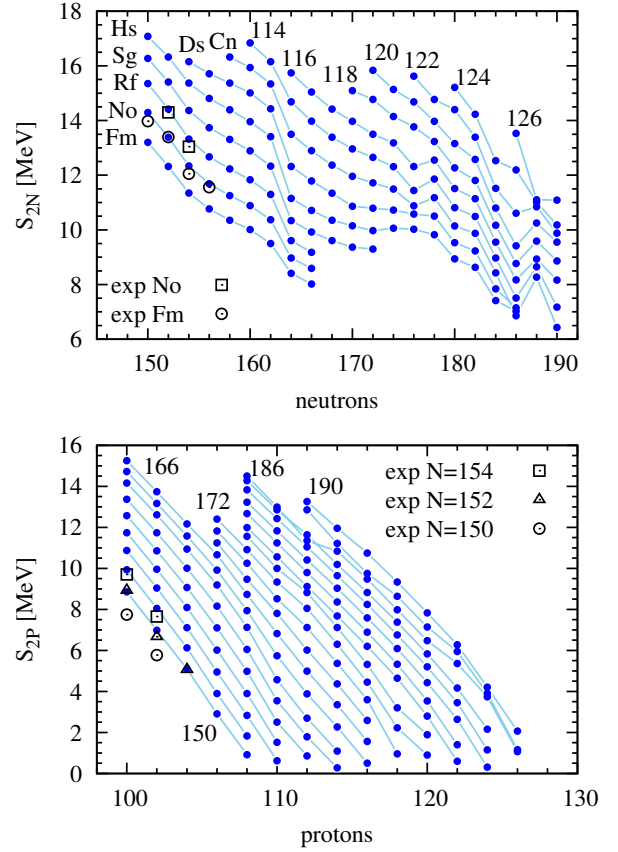}
\caption{\label{fig:S2n}  (Color online) Top panel: Two neutron  separation energies as a function of the neutron number. Lower panel:
Two proton separation energies as a function of the proton number. The lines represent isotonic chains, the one
at the lower left corner  corresponds to N=150, and the one closer to the upper right corner to A=190.  Experimental data are taken from Ref. \cite{aud03b}.}
\end{figure}
\end{center}

   In Fig.~\ref{fig:S2n}, upper panel, we present the 2-neutron separation energy as a function of the neutron number (see
   also column 11 of Table \ref{TAB1}). 
As expected we obtain a decreasing behavior of $S_{2N}$ with increasing neutron number since we get closer to the
neutron drip line. The  smooth  decline of  $S_{2N}$ is only disturbed at the neutron numbers  corresponding to the
single particle shell gaps 162, 184,  etc.  The two-proton separation energies are shown, in the lower panel of 
Fig.~\ref{fig:S2n}, as a function of the proton number (see also column 12 of Table \ref{TAB1}).  The general behavior of decreasing $S_{2P}$ for a given
isotonic chain with growing Z illustrates the fact that we are getting closer to  the proton drip line.  The fact that the 
$S_{2P}$ presents less structure than the $S_{2N}$ is obviously related with the fact that, in the region of interest, the shell gaps  in Fig.~\ref{fig:spe} are smaller for protons than for neutrons.  Fig.~\ref{fig:S2n}  also includes the available experimental values for some Fm and No isotopes.  We observe that in the case of $S_{2N}$ the theoretical values are a  around 1 MeV smaller  as compared to experiment while in the case of $S_{2P}$ they are around 1 MeV larger. In both cases, however,
the trend is correctly described.  This is the  well known binding energy drift  that takes place with the D1S parametrization of the Gogny force for most  isotopic chains. To correct this drift Chappert et al. \cite{D1N} have recently done a new parametrization of the Gogny force, the D1N,  which reduces considerably the drift but otherwise keeps the quality of the D1S parametrization. In Table~\ref{tab:D1SNM} we have included calculations with the D1N parametrization and we observe a considerable improvement in the agreement. We also include in the table the results of calculations with the D1M parametrization of the Gogny force obtained by including beyond mean field effects in the fit \cite{D1M}, which also do not present the mentioned drift.
\begin{table}
\caption{\label{tab:D1SNM}Two-nucleon separation energies and  $Q_{\alpha}$ in Fm isotopes calculated with various Gogny forces compared with experimental data \cite{aud03b} (units are MeV).}
\begin{tabular}{lldddd}

\hline
& $N$    &  D1S &   D1N  &  D1M   &      EXP\\
\hline
\multirow{4}{*}{$S_{2N}$}&150 &  13.20 & 13.86 &14.00  &	   13.98  \\
                         &152 &  12.32 & 13.01 &13.14  &	   13.40  \\
                         &154 &  11.35 & 12.13 &12.23  &	   12.06  \\
                         &156 &  10.77 & 11.64 &11.73  &	   11.56  \\
\hline
\multirow{4}{*}{$S_{2P}$}&150 &	 8.86  &  7.80 &  7.62   &    7.74  \\
                         &152 &	 9.93  &  8.79 &  8.59   &    8.93  \\
                         &154 &  10.87 &  9.67 &  9.47   &    9.71  \\
                         &156 &  11.74 & 10.50 & 10.31   &   10.43  \\
\hline
\multirow{4}{*}{$Q_{\alpha}$}&150 &   7.35 &	 7.66  &  7.70  &     7.56\\
                             &152 &   7.12 &	 7.49  &  7.54  &     7.15\\
                             &154 &   7.02 &	 7.38  &  7.48  &     7.31\\
                             &156 &   6.66 &	 6.99  &  7.10  &     7.03\\
\hline
\end{tabular}
\end{table}

The D1N and D1M parametrizations also improve  slightly the agreement of $Q_\alpha$ with the experiment (see Table \ref{tab:D1SNM}), but this effect is much less pronounced than in the case of the separation energies as the influence of protons and neutrons cancel each other.


\begin{center}
\begin{figure*}
\includegraphics[angle=-90,scale=0.65]{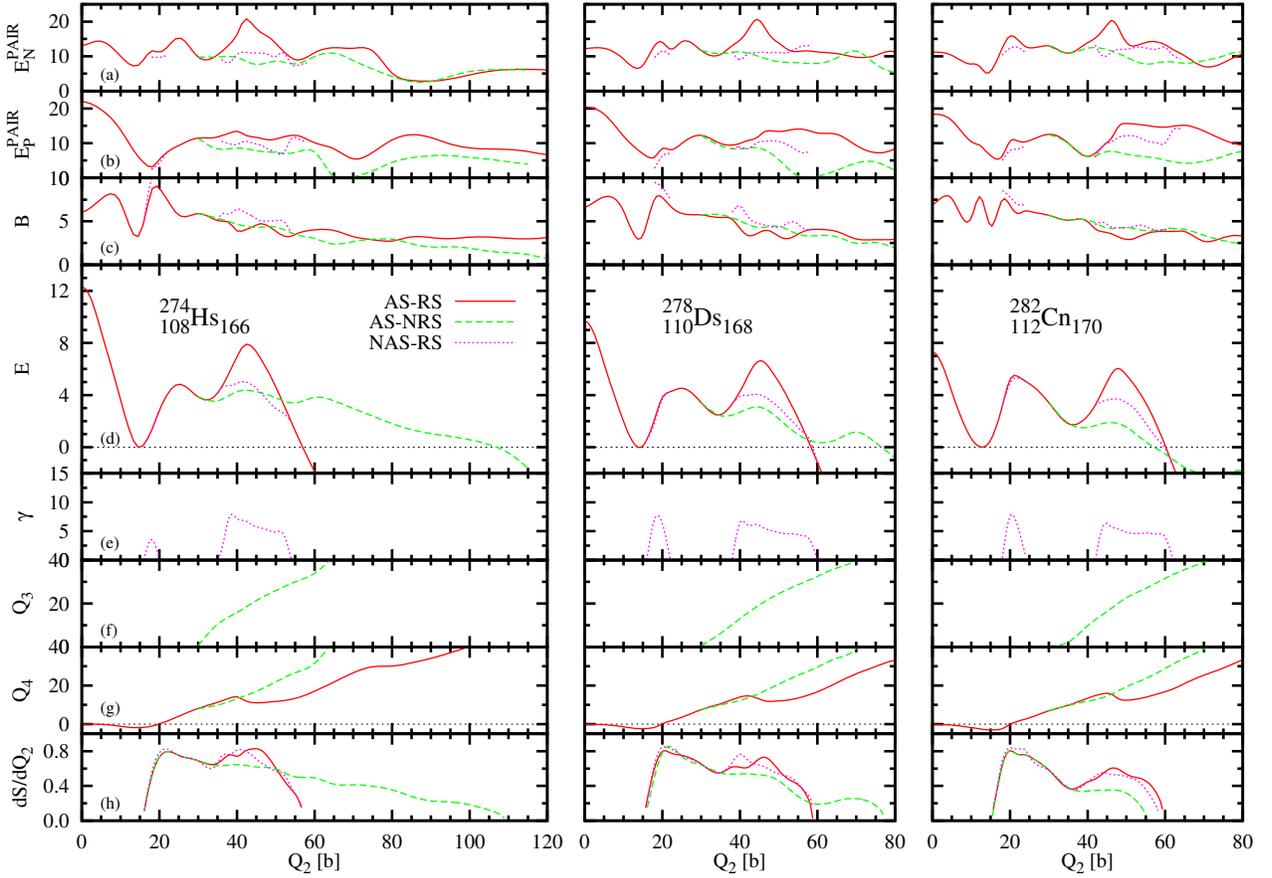}
\caption{  (Color online) Behavior of several magnitudes for the  nuclei $^{274}$Hs,  $^{278}$Ds and $^{282}$Cn along different fission paths. Neutron and proton pairing energies $E_P^N$ (a) and $E_P^P$ (b) are in MeV, (c) inertia parameter $B$ in $10^{-6} b^{-2} MeV^{-1}$, (d) HFB potential energy in MeV, (e) triaxial deformation parameter $\gamma$ in degrees, (f) octupole moment $Q_3$ in $b^{3/2}$, (g) hexadecapole moment $Q_4$ in $b^2$, (h) sub-integral function of action integral $dS/dQ_2$ in  $b^{-1}$.}
\label{8panel3}
\end{figure*}
\end{center}


\section{Spontaneous fission}


\subsection{Symmetry breaking effects in the fission paths.}\label{Sect:IVA}
  
   The large size basis used in these calculations and the large number of nuclei studied prevents to make a systematic study with totally unrestricted symmetry-breaking wave functions.  To clarify the effects of these restrictions we show in Fig.~\ref{8panel3}, as an example,  the behavior of several relevant quantities  along the fission path for the nuclei 
   $^{274}$Hs (left panel),  $^{278}$Ds (middle panel) and $^{282}$Cn (right panel) in different approximations.  
   The panels (d),   in the middle row of Fig.~\ref{8panel3},  display three different fission paths for each nucleus corresponding to the following constraints  on the wave functions:
1.- The Axially Symmetric ($\gamma =0 $) and Reflexion  Symmetric ($Q_3 =0$)  fission path
(continuous line), which we shall call in the following AS-RS path. 2.- The Axially Symmetric and Non-Reflexion
Symmetric ($Q_3 \ne 0$) path (long-dashed lines), in the following AS-NRS path. 3.- The Non-Axially Symmetric 
($\gamma \ne 0$) and Reflexion Symmetric path (short-dashed lines), in the following NAS-RS path.

 In the ground state these nuclei are well quadrupole deformed, $Q_2 \approx15$ b,  see panels (d), and in the way to fission they have to tunnel through a   barrier of several MeV height. The barriers represent the potential energy
needed to deform the nucleus. It is related,  therefore, with the single particle levels around the Fermi surface 
available at the corresponding deformation along the path.  For the three nuclei the paths in the AS-RS case present two hump barriers where the height of the second barrier decreases with the mass number while the first one remains more or less constant for $^{274}$Hs,  $^{278}$Ds and increases for   $^{282}$Cn.  The width of the barrier in this approach is also similar for the three nuclei.    The origin of the two humps can be easily understood looking at the single particle energies of 
Fig.~\ref{fig:spe}. If we follow the neutron Fermi level  to the prolate side we find two regions with a clearly developed low level density.  The first one  at $Q_2\approx 15$ b with the shell gaps N=162 and Z=108 corresponds to the ground state minimum. On the way from this point to larger deformations we find a larger level density region which corresponds to the energy increase of the first barrier.  Behind that we arrive at the second region at $Q_2\approx 32$ b,  which corresponds to the super deformed minimum. At even larger deformations one finds a high level density region that provides the  energy rise  of the second barrier. Finally above $40$ b,  we observe intruder states of a high lying $\nu$i$\frac{13}{2}$ orbital which could be an indication of the scission point. Incidentally, since the neutron shell gap is at N=162 and the proton one at Z=112   the super-deformed minimum is  deeper in $^{282}$Cn than in the lighter nuclei displayed in Fig.~\ref{8panel3}.
  
  This behavior changes remarkably  in the AS-NRS approach.  One can follow in panels (f) the portion of the paths where a
lower solution with  $Q_3\neq 0$ is found.
 As one can see in panels (f) the octupole degree of freedom plays an important role for quadrupole deformations  starting around $Q_2\approx 30-35$ b, i.e., close to  the super deformed minima. 
  As a matter of fact in this channel the second hump of the barriers diminishes strongly and the paths in this region look like the continuation of the first hump. The  AS-NRS path reaches up to large deformations,   $Q_2\approx 108$ b for $^{274}$Hs, $Q_2\approx 77$ b for $^{278}$Ds and decreases to $Q_2\approx 67$ b for $^{282}$Cn.

  We can understand the onset of octupole deformation looking again at  Fig.~\ref{fig:spe}.  Here we observe that  the $K=1/2$ and $K=3/2$  levels of the $\nu$k$_{\frac{17}{2}}$ shell cross the Fermi surface around 26 b, while  the $K=5/2$ level crosses around 35 b. Interestingly, at these $Q_2$ values the $K=5/2$ level of the $\nu$h$\frac{11}{2}$  orbit also approaches the Fermi surface. In  Fig.~\ref{fig:spe} and at zero deformations the $\nu$k$_{\frac{17}{2}}$ subshell  lies at about  0.4 MeV   and the   $\nu$h$\frac{11}{2}$ at about $-10.4$ MeV. The  k$_{\frac{17}{2}}$ and  h$\frac{11}{2}$ subshells  interact
strongly through the $\Delta L =\Delta J = 3$, i.e., the octupole interaction. That means if we allow for
reflexion symmetry breaking  we can increase the quadrupole deformation  at 
a lower energy cost \cite{naz92}. We can observe in Fig.~\ref{8panel3} that around these values the AS-NRS fission paths get lower in energy  than the AS-RS ones.  In the (g) panels we can follow the behavior of the
hexadecapole moment along the fission path. In the ground state it is close to zero in both approximations and from this point
on it grows linearly with $Q_2$ in the AS-NRS approach. In the AS-RS approach, however,  first it increases linearly up to the scission
point where  a kink is observed followed by a linear increase.

  In the NAS-RS path we can observe the effect  of the triaxial shapes along the fission paths. In panels (e) we can see the
two  portions of the trajectory  where  triaxial solutions are found.  The first one,  close to the ground state, spans a smaller 
interval of $Q_2$ values than the second one and  does not have a large impact  on the energy. The second one, as in the AS-NRS case, is relevant around the second hump causing a significant lowering of its height, i.e., in this case we still have to deal with two humped barriers. The width of the barrier, at variance with the AS-NRS case, is more or less like the AS-RS one. 

Though the shape of the barriers is very relevant to calculate lifetimes one has to consider, however, that other quantities entering in the corresponding formula,   Eqs.~(\ref{TSF}) and (\ref{ACTION}), do play an important role. A relevant parameter is the collective quadrupole inertia $B(Q_2)$, since mass parameters are strongly influenced by pairing correlations and these itself by the single particle level density which we expect to vary along the different fission paths. In panels (a) and (b) of Fig.~\ref{8panel3} we display the neutron and protons pairing energies, respectively. We indeed observe big differences in both of them along the portions of the paths where the symmetry breaking takes place. In particular we observe that the
AS-RS solutions provide always the largest pairing energies.  As we can observe in panels (c), where the $B(Q_2)$'s are plotted  in the different approaches, in the relevant parts the AS-RS provides the smaller masses followed
by the AS-NRS ones.  This implies, that not necessarily the smallest fission barriers provide the shortest lifetimes \cite{bar81}.  Since the action
$S$ in Eq.~(\ref{ACTION}) can be seen as a line integral of the function $dS(Q_2)/dQ_2 = (2B(Q_2)(V(Q_2)-E_0)^{1/2}$, the area
of the surface delimited by this line and the X-axis provides the value of the action $S$.  In panel (h) we display $dS(Q_2)/dQ_2$
in the three approaches. For the nucleus $^{274}$Hs we find that though the NAS-RS fission path has a smaller fission 
barrier than the AS-RS, the actions $S$ for both approaches are very close. The actual values of $S$ are 26.69 (AS-RS) and 26.49 (NAS-RS), while  in the  AS-NRS case we obtain a much larger value, namely 41.66. In the nucleus  $^{278}$Ds, the
fission paths alone would predict that the AS-NRS and the NAS-RS approaches would provide much shorter lifetimes than the AS-RS one. However, in panel (h) one finds that the three areas look rather similar.  Actually the precise numbers 25.88 (AS-RS), 26.97 (AS-NRS) and 26.49 (NAS-RS) show this to be the case. Lastly, for $^{282}$Cn, the prediction of the fission paths
is more or less in accordance  with the one of panel (h) and  the actual numbers  23.34 (AS-RS), 18.32 (AS-NRS) and 22.52 (NAS-RS) corroborate that.   We can conclude that the restriction to axially symmetric paths is, in general,  a good approximation, though as we we will see later one can find some exceptions.

The low and short NRS barrier in nuclei with $N\ge 170$ makes the most probable fission through octupole deformed shapes. In these nuclei we expect to find asymmetric mass distribution of fission fragments \cite{poe12}.

\begin{center}
\begin{figure}
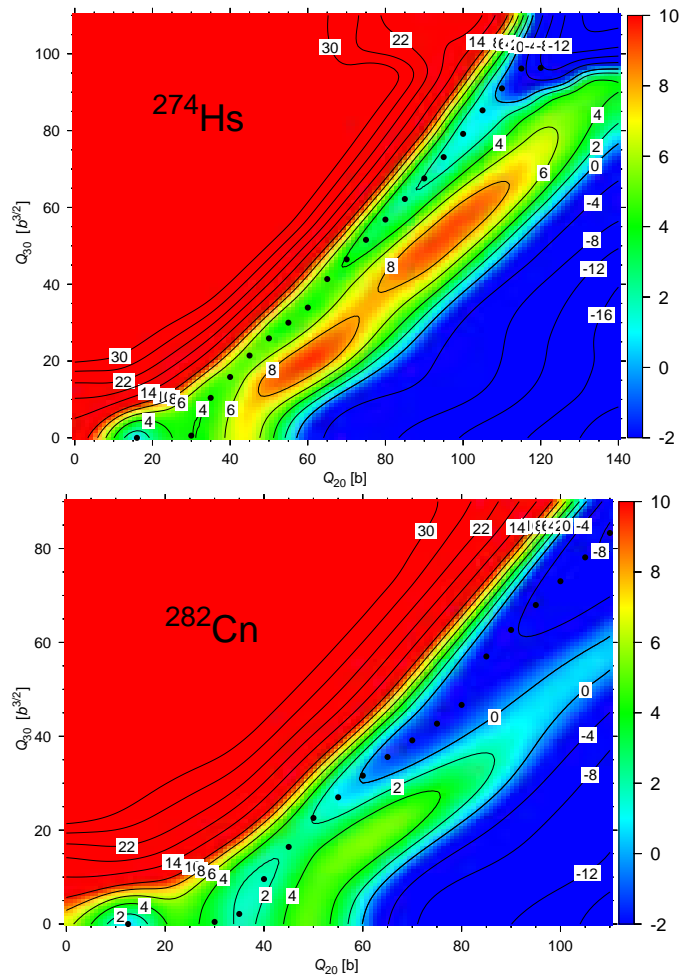

\includegraphics[angle=-90,scale=0.4]{q2-q3_274Hs.ps}
\includegraphics[angle=-90,scale=0.4]{q2-q3_282Cn.ps}
\caption{\label{q2-q3}  (Color online) The PES of $^{270}$Hs as a function of quadrupole and octupole moments.  The energy origin has been set at the energy minimum.}
\end{figure}
\end{center}

 In the one dimensional fission paths plotted in Fig.~\ref{8panel3}  we find crossings between the two paths giving the
 impression that one could switch from one path to the other without further problem. 
 However, if we look in a higher dimensional plot one can see that this is not the case. To illustrate this point we have
 drawn in Fig.~\ref{q2-q3} potential energy contour lines versus the quadrupole, $Q_2$, and the octupole moment, $Q_3$,
 for the nuclei  $^{274}$Hs and $^{282}$Cn. In this figure we can follow the AS-RS and the AS-NRS paths of Fig.~\ref{8panel3} for the respective nucleus. The AS-RS path corresponds to $Q_3=0$ and goes along the X-axis
 and the AS-NRS one goes along the bullets. It is interesting to see how the self-consistent path goes along a valley 
 in both nuclei. We can also see that no alternative paths are present. In the $^{274}$Hs case we find that, at $Q_2=50$ b where both paths seem to cross in  Fig.~\ref{8panel3}(d), in reality both paths are separated by  a 4 to 5 MeV high barrier. 
  
\subsection{Fission barriers in the axially symmetric approaches.}
 
\begin{figure*}
\includegraphics[scale=0.90]{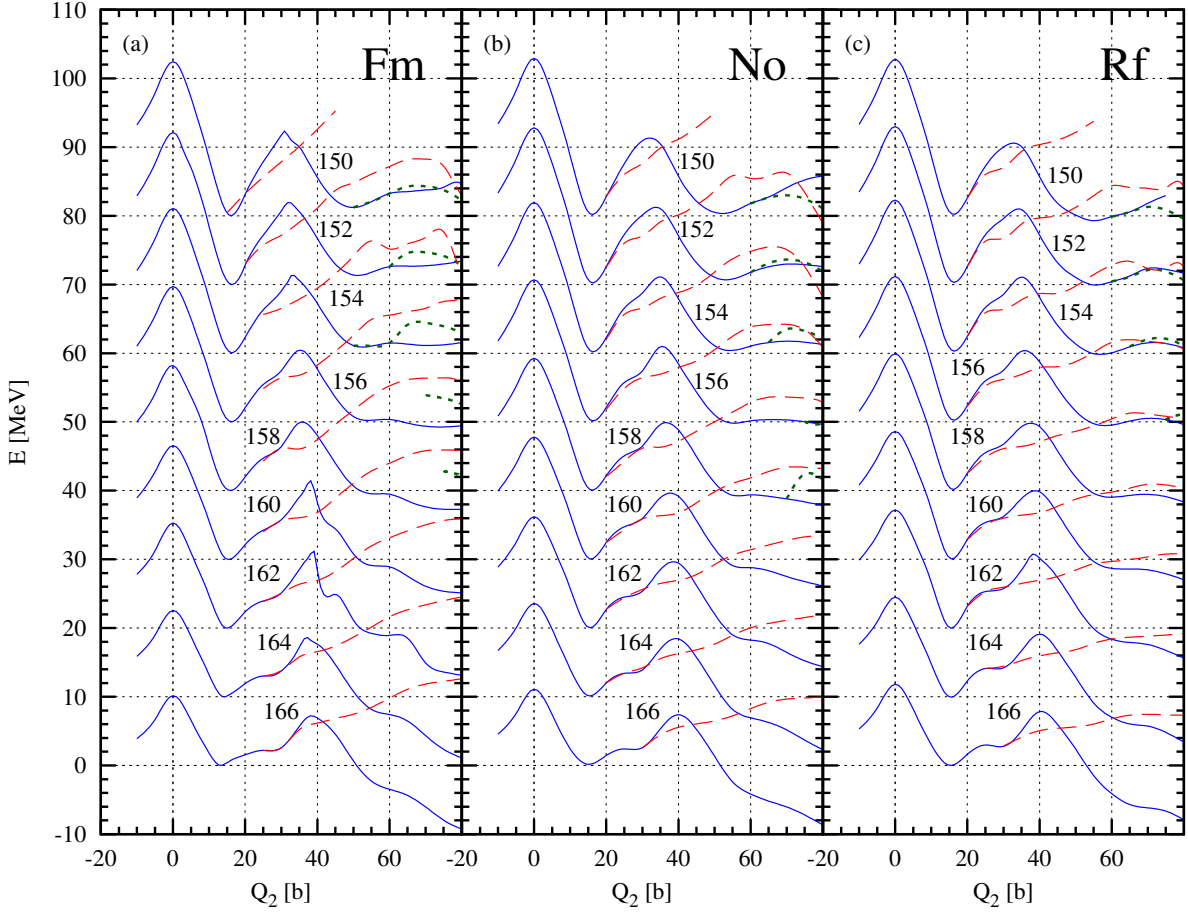}
\caption{\label{fig:Fb100-104}  (Color online) Fission barriers for the nuclei Fm, No and Rf along the AS-RS  (continuous lines) and
the AS-NRS path (dashed lines).  The ground state energy of each isotope has been  set to zero and 
the barriers of the different isotopes have been shifted in 10 MeV in the axis of coordinates
for clarity reasons. }
\end{figure*}

\begin{figure*}
\includegraphics[scale=0.90]{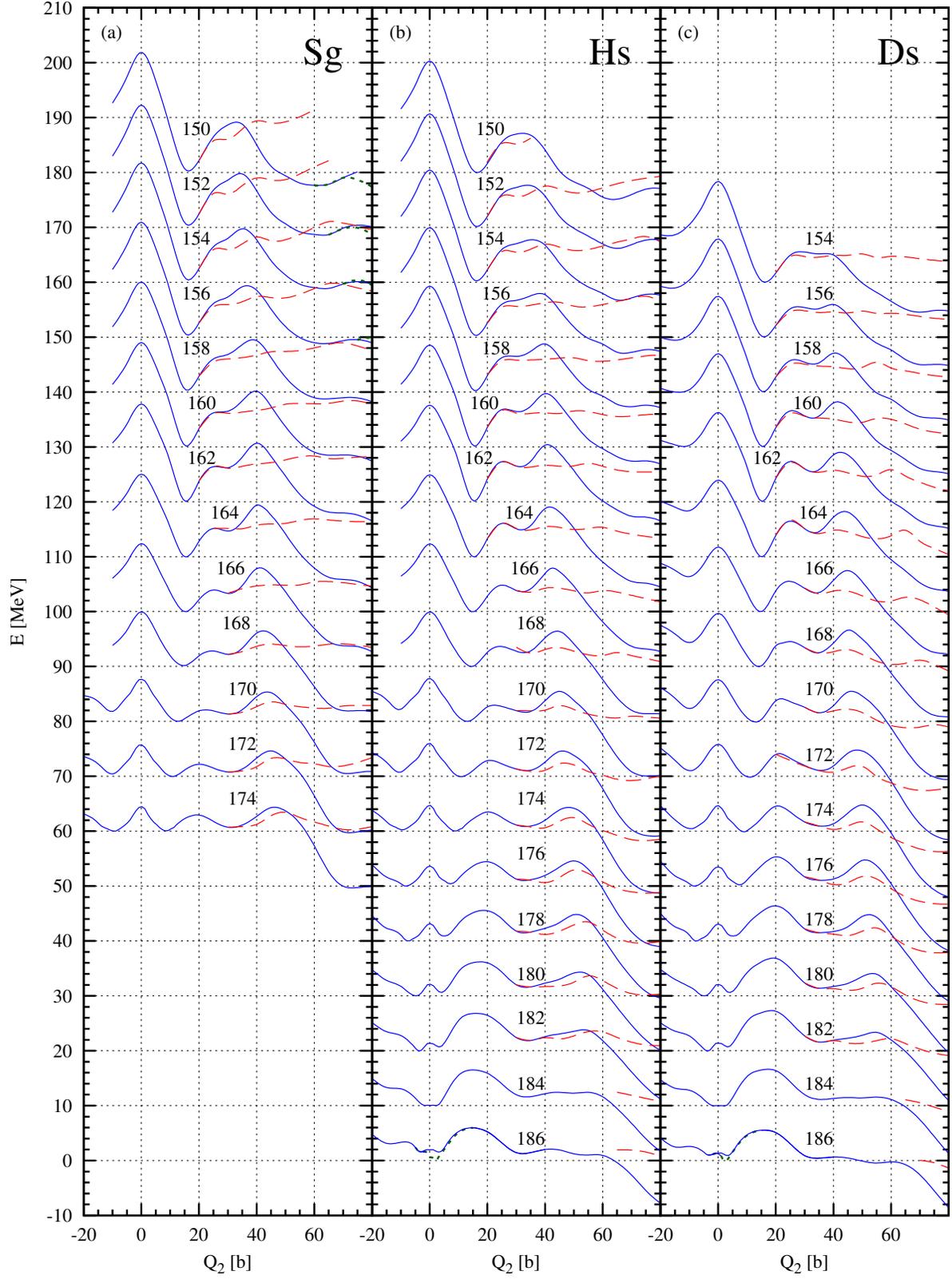}
\caption{\label{fig:Fb106-110} 
(Color online) Fission barriers for the Sg, Hs and Ds isotopes along the AS-RS  (continuous lines) and
the AS-NRS path (dashed lines).  The green short-dashed lines around the ground state for $A > 184$ correspond to NRS solutions. The ground state energy of each isotope has been  set to zero and 
the barriers of the different isotopes have been shifted in 10 MeV in the axis of coordinates
for clarity reasons. }
\end{figure*}

\begin{figure*}
\includegraphics[scale=0.90]{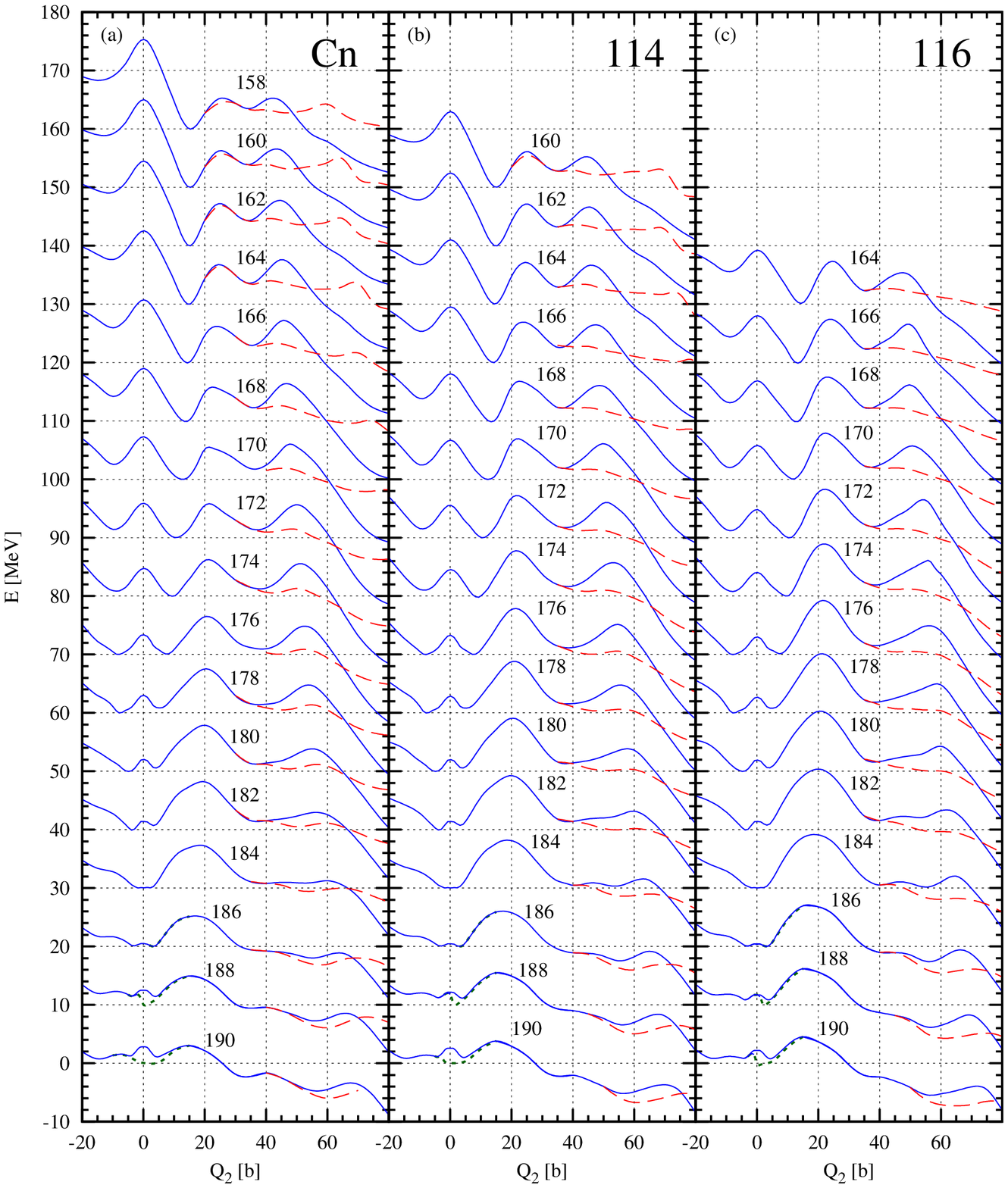}
\caption{\label{fig:Fb112-116}  (Color online) Same as Fig.~\ref{fig:Fb106-110}, but for the Cn, $Z=114$ and 116 isotopes.}
\end{figure*}

\begin{figure*}
\includegraphics[scale=0.90]{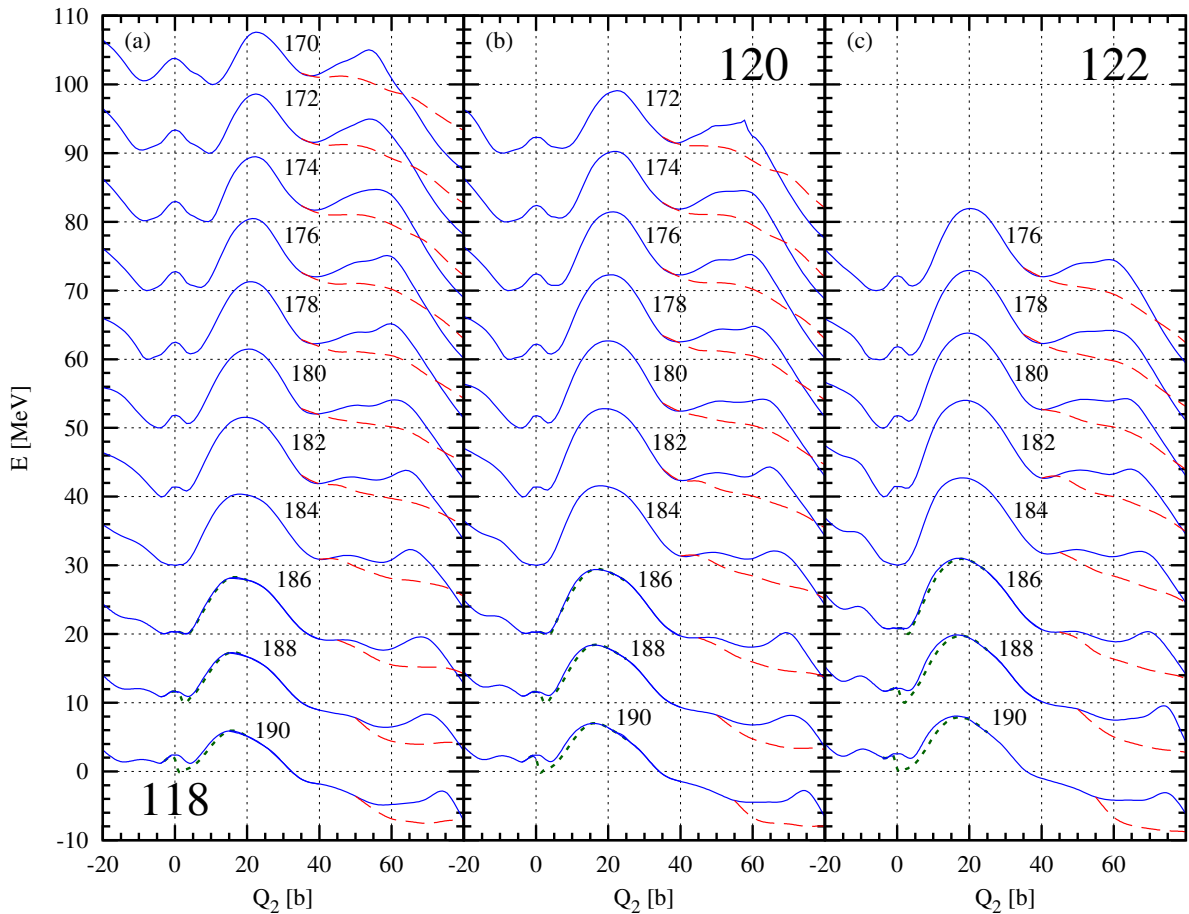}
\caption{\label{fig:Fb118-122}  (Color online) Same as Fig.~\ref{fig:Fb106-110}, but for the $Z=118, 120$ and 122 isotopes.}
\end{figure*}

\begin{figure*}
\includegraphics[scale=0.90]{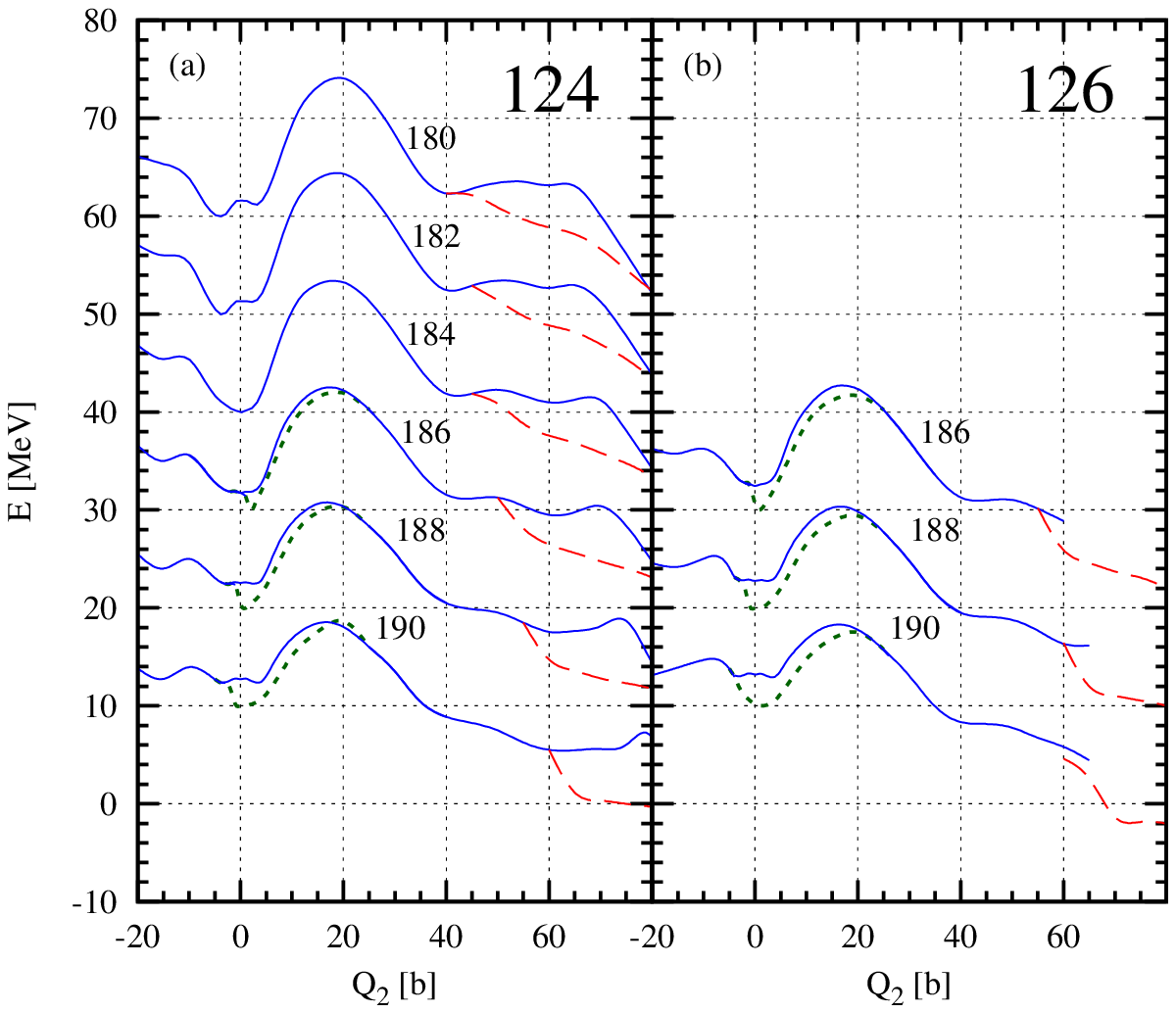}
\caption{\label{fig:Fb124-126}  (Color online) Same as Fig.~\ref{fig:Fb106-110}, but for the $Z=124$ and 126 isotopes.}
\end{figure*}

 In the following to perform a systematic description of the fission barriers of the 160 SHE's, we restrict ourselves to  the axial approximation in  which  we have performed two kinds of calculations, namely the reflexion symmetric, AS-RS, and the non-reflexion symmetric, AS-NRS.    
  All the fission barriers are presented in the series of Figures  \ref{fig:Fb100-104}-\ref{fig:Fb124-126}.

In Fig.~\ref{fig:Fb100-104}
we present the fission barriers for the isotopes of the elements Fm, No and Rf for quadrupole values from
$-20$ b up to $80$ b (continuous line for AS-RS and dashed for AS-NRS). We first discuss the AS-RS results.
In panel (a) we present the Fm results
for neutron number 150 up to 166. All isotopes present a well prolate deformed minimum around 15 b.  In addition in the 
lighter isotopes  a shallow superdeformed (SD) minimum appears around 50 b, at $N=156$  we find a very 
flat minimum and for the heavier isotopes no SD minimum is found. The common characteristic of these 
nuclei is the presence of a big broad barrier. For $N=150$ the barrier is centered at $Q_2=30$ b, 
has a height of about 12 MeV and a width of 18 b.  With
increasing neutron number the center of the barrier shifts to larger deformations and the height diminishes.
For $N=166$ the center is around  $Q_2=38$ b and the height is about 8 MeV.
In the heavier isotopes we find some structure in the first barrier, 
namely the development of a shoulder  around $Q_2= 27$ b with increasing neutron number.  The presence of 
a SD minimum in the lighter isotopes drives the  existence of a second barrier.  Since the minima are 
rather shallow the second barriers are broad but not high.  These properties will contribute in general to a
tendency of shorter lifetimes with increasing neutron number, though the particular behavior must be analyzed 
case by case. 

The fission paths for the No isotopes are shown in  panel (b) for the same neutron numbers
as the Fm case. The structure of the first barriers look roughly like the ones in Fm.  An important point
is the fact that the SD minimum is somewhat lower in energy than in the Fm case and its role is therefore much
less relevant. The results for the Rf isotopes are displayed in panel (c).  The tendency observed in
the No isotopes is reinforced, the SD minimum gets even deeper and the second barrier disappears for most 
isotopes. This fact will provide in general shorter lifetimes for Rf than for No and for No shorter than
for Fm. These are general tendencies but since the lifetimes are very sensitive to small energy differences 
along the fission path, 
to make quantitative predictions calculations involving also collective inertia have to be performed.

Let us now describe the AS-NRS results.  For all isotopes of the three elements, the fission barriers are
much larger along the non-reflexion symmetric paths than along the reflexion symmetric path.
The lifetimes are therefore considerably longer in the AS-NRS path than in the AS-RS 
one. The fission in the AS-NRS  mode is completely impossible.

An exhaustive discussion of the fission barriers of the nuclei $^{254}$Fm, $^{256}$Fm, $^{258}$Fm, 
$^{258}$No, and $^{260}$Rf can be found in Ref.~\cite{werp}. In these nuclei the ``elongated fission" mode can be observed. This mode is connected to octupole deformed fission paths which start at $Q_2>50$ b. We have not observed the ``elongated fission" in heavier nuclei therefore we will not discuss it here.

In Fig.~\ref{fig:Fb106-110} we present the paths for the nuclei of Sg, Hs and Ds elements. In panel (a) we display the results
from $N=150$ up to $N=174$ for the Sg isotopes. These nuclei have a trend somewhat different from the preceding ones:
the SD minimum does not play a relevant role in the fission process since it is always deeper than the ground state and
as a matter of fact if the SD minimum were not that shallow it would be the  ground state for the lightest Sg isotopes.
Furthermore the barriers get flatter, most of them being lower than 10 MeV.
 Apart from this feature the situation
for the isotopes $N=150-166$  is similar to the nuclei Fm, No and Rf, i.e., the ground states are prolate deformed 
($Q_2= 15$ b) with large negative deformation energies and high barriers  at larger $Q_2$ values. For $N=168-174$ the situation changes 
very fast, and  the following properties get 
reinforced as the neutron number increases: the ground states get less deformed, the shoulders  around $Q_2= 30$ b 
get deeper and become  real minima and as a consequence the original barrier becomes a two humped one. The height
of the spherical maximum  decreases and  a coexisting oblate minimum develops 
at $Q_2= -10$ b.  In particular for $N=172$ the oblate and prolate minima are degenerated.
We therefore expect a strong reduction in the  fission lifetime of heavier isotopes as 
compared to the lighter ones.

  In panel (b) the $N=150$ up to $N=186$ Hs isotopes are shown. In the lighter isotopes the fission 
barriers are flatter and  the SD minima  deeper than in the corresponding Sg isotones. The effect of the SD minima on the fission process is much smaller. Furthermore the second fission barrier develops much earlier. Since the single particle energies of Fig.~\ref{fig:spe}
are for $^{270}$Hs we can confront the general discussion of Sect.~\ref{Sect:IVA} with the shape of the fission path in this nucleus.
Aside from these facts
the Hs isotopes from $N= 150$ up to $N= 174$ behave to a large extend like the corresponding isotones of the Sg 
isotopes. For $N=176$ the oblate minimum becomes the ground state and the two humps  of the fission barrier 
are very similar in size and height (about 5 MeV). With increasing number of neutrons,  the prolate minimum shifts towards
 $Q_2=0$ and the spherical minimum becomes the ground state at the $N=184$  shell closure. As a consequence the
two humps separate from each other, the outer one shifting to larger $Q_2$ values and the inner one to smaller ones. At 
 the same time the inner barrier gets bigger and the outer smaller. From this behavior one expects the lifetimes
for fission  to be smaller around neutron number 170-174. 

 The nucleus $^{294}$Hs is octupole deformed in its 
ground state, see Fig.~\ref{fig:beta}.  In the self-consistent $Q_2$-constrained calculations this nucleus remains weakly octupole deformed 
up to $Q_2=10$ b where it turns reflexion symmetric (see short dashed part of the fission path). Since the fission fragments at the scission point are
characteristic of a reflexion symmetric fission we still denote this mode as RS fission. The same situation is found with the $N>184$ isotones for the 
heavier SHE. The paths along $Q_3\ne 0$ close to the ground state are plotted with short dash lines in the corresponding figures. 

The Ds results for the isotopes $N=154$ up to $N=188$ are displayed in panel (c). The main characteristics
of these fission paths are the following:  For the lighter isotopes the fission barriers are flatter 
than for the corresponding isotones in Hs and Sg. For the light and medium mass isotopes, for a given 
isotone number, we find an increase of the first barrier  moving from Sg to Hs and from this to Ds. 
The opposite effect is observed for the  second barrier, in particular for the very
heavy isotopes this barrier disappears at the highest neutron number studied. 
  
Concerning the AS-NRS results of Fig.~\ref{fig:Fb106-110} and for the Sg isotopes,
we find that though the AS-NRS fission barriers are smaller than the
AS-RS ones for medium $Q_2$ values, for larger  ones   they are much higher and as 
a result the non-reflexion symmetric fission paths are not favored as compared  with the 
reflexion symmetric ones.  For the Hs  isotopes the same behavior as 
the Sg isotopes is observed: for lighter isotopes up-bending tails of the fission paths make the
AS-NRS paths very unfavored,  but for medium-heavy N values the tails bend down and though the path is
longer the second barrier is smaller for the AS-NRS path than for the AS-RS one,
i.e., around $N=170$ and above  the AS-NRS path may compete with the AS-RS one (cf. Fig. \ref{q2-q3}). For the heaviest Hs 
isotopes in the AS-RS approach the second barrier decreases considerably and the AS-NRS barrier becomes
longer than the other one.  For the Ds isotopes the AS-NRS path is even more favorable because the down-bending
tendency gets reinforced and we have some AS-NRS paths which are clearly favored, for example for the nuclei with
$N=170-176$. For larger N values the vanishing of the second barrier in the AS-RS case again favors this 
approach. Notice that in general octupole effects set in for larger $Q_2$ values as the neutron number increases.

In Fig.~\ref{fig:Fb112-116} we present the fission paths for the isotopes of the elements Cn and $Z=114$
and 116. As before we first discuss the AS-RS paths. In panel (a), where the isotopes $N=160-188$ of Cn 
 are shown, we observe that compared with the corresponding isotones 
discussed before, the second hump of the barriers is lower and that the slopes of the tails of the fission paths
are more pronounced.  These facts point  to shorter lifetimes of the Cn 
isotopes as compared with lighter isotones.
  In  panel (b) of the figure the $Z=114$ isotopes are shown.  The main difference with respect to the 
Cn isotopes is the increase of the first hump and the decrease of the second one. In the isotopes with 
 one hump barriers these are higher and a bit broader than for the corresponding isotones in Cn. Altogether, it seems that, in general, the lifetimes 
of the $Z=114$ isotopes will be somewhat longer than the one at the corresponding   $Z=112$ isotones. In panel
(c) we display the $Z=116$ isotopes. Here the same trend as in the previous nuclei is observed: a reinforcement
of the tendency to increase the first hump of the double humped barriers and in the case of only one hump an increase
of this.

  Concerning the AS-NRS results for Cn and the  $Z=114$ and 116 elements, we find that the onset of octupolarity
is energetically favored after the level crossing of  the ``higher shells"  and with the
exception of the lightest isotopes the barriers are much smaller in the AS-NRS path.  This is due to the disappearance of the
second hump of the barrier, i.e., in the AS-NRS path we have only one-humped barriers. We expect therefore a shortening 
of the fission lifetimes along these paths.

 In  Fig.~\ref{fig:Fb118-122} the results for the  $Z=118, 120$ and 122 isotopes is shown. The same tendency as 
before is observed in the AS-RS calculations, larger first barriers as $Z$ increases. The role of the octupole degree
of freedom is also relevant and all AS-NRS fission paths do have smaller fission barriers. The same comments do also 
apply to the $Z=124$ and 126 isotopes in Fig.~\ref{fig:Fb124-126}. 

\section{Half-lives of SHE's}
\label{H-L_sect}

 \begin{figure*}
\includegraphics[scale=0.75]{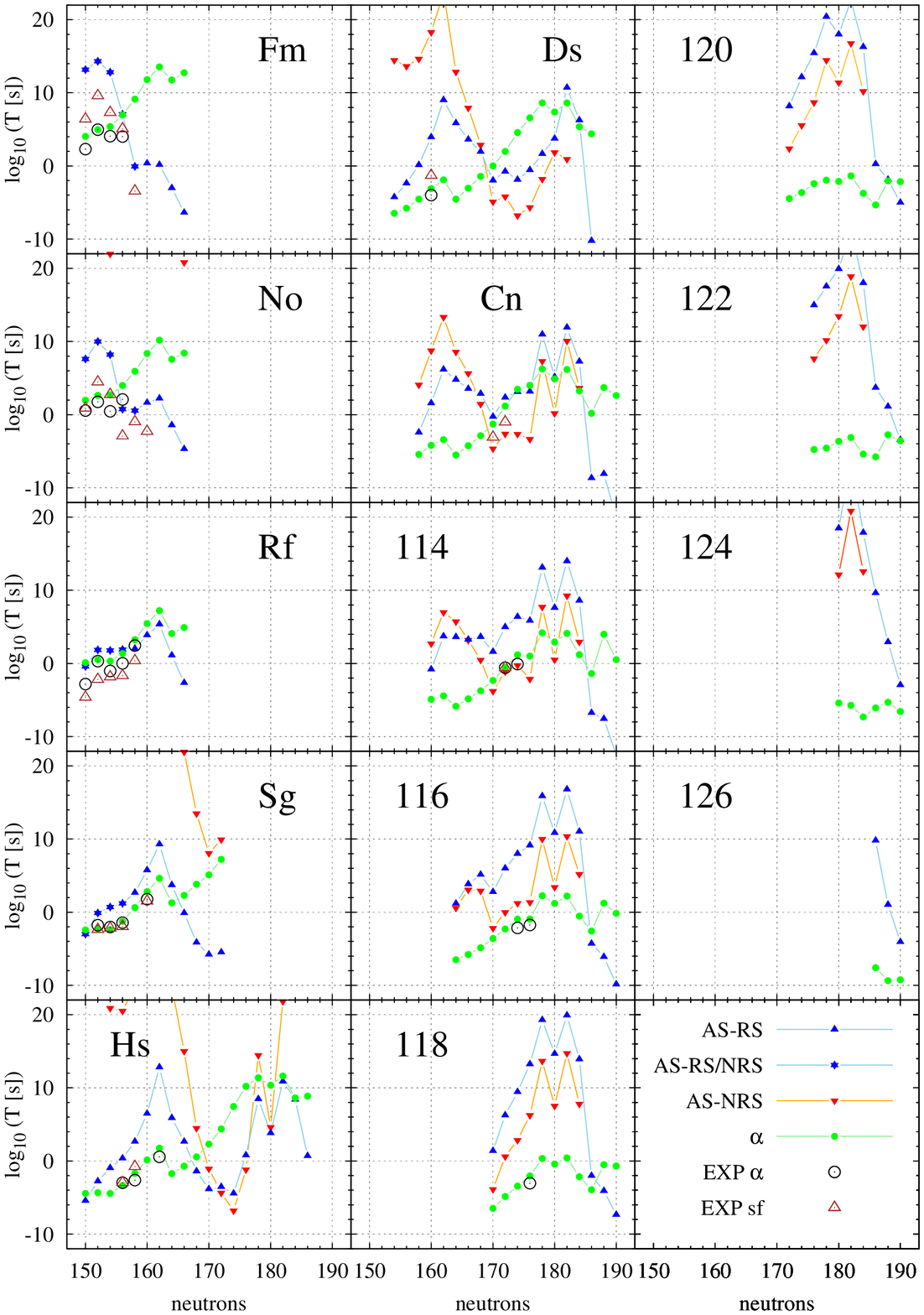}
\caption{\label{fig:T1/2}  (Color online) The spontaneous fission and the $\alpha$ emission half-lives in SHE's. Experimental data are taken from Refs. \cite{oga05,oga07,aud03a}}
\end{figure*}

\begin{figure}
\includegraphics[scale=0.35, angle=270]{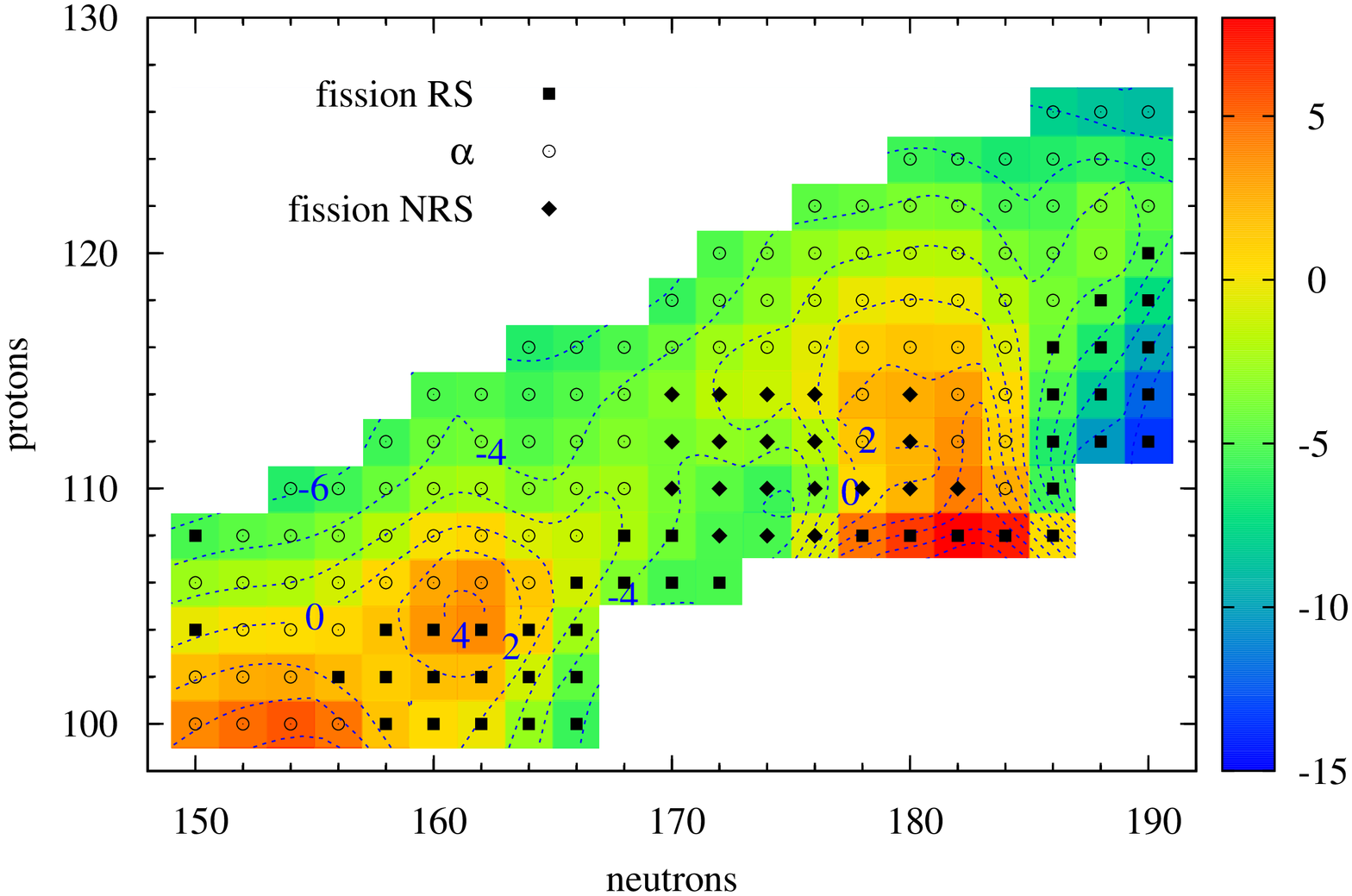}
\caption{\label{fig:T1/2_main} (Color online)  The dominant decay modes of SHE's. Logarithm of the shortest half-lives (in s) is also indicated. }
\end{figure}

One of the key issues in the theoretical description is the prediction of the decay modes and the half-lives of the SHE's. The agreement between the experimental data and the theoretical predictions states not only the quality of both of them but  it may be  also a criterion for the identification of a particular isotope synthesized in the corresponding fusion reaction.
Moreover, since the contemporary experimental techniques do not allow the detection of nuclei with half-lives shorter than   $T = 10\; \mu$s one must estimate which isotopes can survive long enough to be detected and the way in which they disintegrates. Therefore, the half-lives of the two main competing processes, spontaneous fission and $\alpha$ emission, should be evaluated. The shortest half-life determines the dominant decay channel and the total half-life. If the branching ratio between two modes is equal to 50~\% the logarithm of the total half-life would be smaller than the shorter partial half-life by not more than 0.3. Differences in half-lives between two modes of one order of magnitude lead to a logarithm of the total half-life only 0.05 smaller than the logarithm of the half-life of the fastest decay.

The half-lives for $\alpha$ decay and  spontaneous fission calculated in the HFB theory are collected in the last two columns of Table \ref{TAB1}.  For an easier analysis these data are also presented in Fig.~\ref{fig:T1/2} where the isotopic chain of each element is shown in a separate panel. In some nuclei two fission half-lives can be calculated along  paths leading to distinct fragment mass asymmetry. Both solutions are depicted in Fig.~\ref{fig:T1/2} and discussed below, however, in Table I only the shorter half-life of the dominant mode is given. The available experimental data  are also plotted in Fig.~\ref{fig:T1/2}.

A first look into  the panels of Fig.~\ref{fig:T1/2} reveals similar tendencies for the different isotopic chains. It is easy to distinguish the intervals of neutron number where common features are characteristic for many elements despite of differences in the absolute values. Therefore we will discuss our results collected in groups of similar neutron number starting from the lighter ones.

 As we have seen in the discussion of the precedent section NRS effects influence  the fission  paths,  and thereby the fission half-lives, in three $Q_2$ regions: 1.- For small $Q_2$ values, $Q_2 < 20$ b, they affect nuclei with octupole deformed ground states, i.e. nuclei with $N >184$. 2.-
 Starting at medium  ($Q_2 > 20$ b ) and extending up to large $Q_2$ values, these effects are present in all nuclei. 3.- For larger  $Q_2$ values, $Q_2 > 50$ b,  they appear in the light isotopes of the elements Fm, No, Rf and Sg.   According to these effects  we are using three symbols in Fig.~\ref{fig:T1/2}  for the fission half-lives depending on the paths used in the calculations~:
 AS-RS (triangle-up) includes the genuine AS-RS ones  plus those with octupole effects close to the ground state. The reason to include the latter under a ``AS-RS'' denomination is that at $Q_2 \approx 20$ b the nucleus takes  AS-RS shape and the fission takes place exactly in the same way as in the pure AS-RS case with symmetric fragment mass distribution.  AS-NRS (triangle-down)  denotes the ``real" NRS paths leading to fission with fragments of different masses, which correspond to the long dashed lines in the fission barrier plots. AS-RS/NRS (star) labels the paths where the first part is AS-RS and only for  $Q_2 > 50$ b  one follows the  NAS branch.

\subsection{The region $150\le N\le 162$}
The first characteristic region covers the lighter isotopes with $150\le N\le 162$, clearly delimited by the $N=162$ deformed ``shell closure''.   In this range of $N$ the half-lives for $\alpha$ emission increase monotonically with the neutron number in all elements from Fm to $Z=114$. It can be also noticed that an increase of the proton number by two units leads to the decrease of the $\alpha$ half-lives by two or three orders of magnitude. As we will also see for the other regions, this tendency is a direct consequence of the calculated values of $Q_\alpha$ shown in Fig.~\ref{fig:Qalpha} and  is consistent with well-known properties of $\alpha$ decay in heavy nuclei. 

The three lightest Fm and No isotopes are a specific group of nuclei  in which the HFB calculations predict  relatively long fission half-lives. The small fission probability is the consequence of the shape of the fission barrier in these nuclei which extends to  large quadrupole deformation. In the AS-RS path the second barrier extends up to 120 b providing very long half-lives. A somewhat shorter half-life  is obtained in the AS-NRS approach where the barrier  is constructed from two humps: the first a RS and the second a NRS one starting at $Q_2 \approx 60$ b (Fig.~\ref{fig:Fb100-104}).
In the slightly heavier isotopes the second NRS barrier also can be noticed in the PES (Fig.~\ref{fig:Fb106-110}), but in contrast to the group of lighter Fm and No nuclei the second, super deformed  minimum has an energy below the ground state.  Thus the second hump does not affect the barrier tunneling but it  governs the fragment mass asymmetry. In these nuclei the reduction of the width of the barrier leads to a decrease of the half-lives by a few orders of magnitude what can be seen in the Fm and No isotopes with $N\ge 156$ as well as the Rf and the Sg isotopes.
  All nuclei with the second NRS hump leading to the so-called ``elongated fission" \cite{werp} (i.e. Fm and No with $N\le 158$ as well as Rf and Sg with $N\le 156$) are marked in Fig.~\ref{fig:T1/2} by the blue stars.

In the heavier isotopes of the  $150\le N\le 162$ interval, where only  the single RS barrier remains, one observes  a rise of the fission half-lives with the neutron number. The slope of this trend changes from an almost flat dependence in Fm to a very steep one in Sg, Ds, Cn, and $Z=114$.  This trend is caused by a broadening of the barrier that eventually becomes a
two humped one (Figs.~\ref{fig:Fb100-104}, \ref{fig:Fb106-110}, \ref{fig:Fb112-116}).

The local maximum found  in all elements in the partial half-lives at $N=162$ with respect to $\alpha$ and fission decays  indicates  the special character of this neutron number in the chart of nuclides, see also Fig.~\ref{fig:spe}. These isotones are the most stable in the close vicinity, hence, in many papers  $N=162$ is called ``deformed magic number". This name stresses the significant difference from the classical magic nuclei which are spherical in their ground states.

Many experimental data coming from the ``cold fusion" reactions are available in the region of isotopes with  $N\le 162$ \cite{aud03a, oga05}. The agreement of the $\alpha$ decay half-lives with the theoretical predictions is noticeable although some discrepancies are observed. Most calculated fission  half-lives  overestimate the experimental data. The agreement would be  better if  triaxial effects  were taken into account in the saddle point of the first barrier. The consideration of the $\gamma$-deformation in the calculations of nuclei in the Fm region reduces the barrier heights by around 2 MeV decreasing thereby the theoretical fission half-lives by around 2 orders of magnitude \cite{werp}. At variance with the discussion of Sect.~\ref{Sect:IVA} in these nuclei this reduction is not fully compensated by an increase of the inertia parameters. Therefore discrepancies between the theory and the experiment diminishes considerably. 
It should be pointed out that the fastest decay is properly predicted in all  cases. Furthermore,  when the experimental data for $\alpha$ emission and fission provide comparable decay probability for both processes the theoretical  predictions are also similar for both partial half-lives.  Almost all nuclei in this region do have  half-lives long enough to be considered as of experimental interest. 

\subsection{The region $164\le N\le 178$}
The next region covers nuclei with neutron number from $N=164$ to $N=178$. At $N=164$  a kink of 2-4 orders of magnitude in the $\alpha$ decay half-lives can be observed corresponding to a local maximum in $Q_\alpha$, see Fig.~\ref{fig:Qalpha}. The increase of the neutron number for a given isotope causes a linear growth in the  $\alpha$ half-life following the tendency already observed  in the lighter nuclei.  The half-life rise up to the neutron number $N=178$ where the locally longest living isotopes are found, see also Fig.~\ref{fig:Qalpha}. They reach values from $T=0.1$ s in $Z=118$ to  $T=10^{11}$ s in Hs. Again, the half-lives calculated for heavier isotones are smaller and, consequently, the $\alpha$ emission becomes the most probable decay channel in the proton rich nuclei.

The AS-RS fission half-lives  for nuclei with $164\le N\le178$ behave completely different than those for $150\le N\le 162$. After passing the maximum at $N=162$ they decrease up to the local minimum at $N=170$. The drop is very steep for the lighter elements (Rf, Sg, Hs) and gentler for the heavier ones (Ds, Cn), whereas for the isotopes of $Z=114$ almost no change is observed. This behavior can be explained by the decreasing potential energy along almost the whole barrier,  associated with a diminishing of the saddle point energy and the clear development  of the second minimum (Figs.~\ref{fig:Fb100-104}, \ref{fig:Fb106-110}, \ref{fig:Fb112-116}). The inverse trend of increasing the potential energy at small deformations is noticed in heavier elements, starting from Cn  (Fig.~\ref{fig:Fb112-116}).
The first RS hump of the barrier grows up with increasing mass of a nucleus. In  the heavy isotopes with $A\ge 280$ it is higher than the second RS barrier.

The half-lives calculated for $162\le N\le 170$ along the NRS path diminish more rapidly with the increase on $N$ than in the RS mode. This is induced by the changes that take place in the  NRS barrier, namely the fast decrease of its height and, more important, the narrowing of its width, see  Figs.~\ref{fig:Fb106-110}, \ref{fig:Fb112-116}.
In nuclei with  $N\ge170$ the NRS fission barrier allows to avoid the second hump of the RS barrier making the asymmetric fission the most probable mode.  Very short fission half-lives, even below $T=1$ ms, can be found in the Hs, Ds, Cn and  $Z=114$ isotopes, where  the NRS fission is the dominant decay mode being even faster than the  $\alpha$ emission. In  $Z=114$ the NRS fission has half-lives comparable to the $\alpha$ decay.

 The available  experimental data around $N=172$ obtained from the ``hot fusion" experiments \cite{oga05,gat11}, fit perfectly to the theoretical predictions. In Cn two spontaneous fission half-lives correspond to the prediction of the NRS fission mode. In $Z=114, N=172$ the observed 50\% branching ratio is  very well reproduced. In the isotope with two more neutrons the detected $\alpha$ decay  is predicted with only slightly longer half-life than in the dominant fission channel.
 Finally, in $Z=116$ and $Z=118$ the dominant $\alpha$ radioactivity is properly predicted by the theoretical analysis with a good estimation of the half-lives.

While approaching  $N=178$ the fission half-lives increase with a slope which grows with the proton number. This effect is governed by two trends observed in the evolution of the PES's.  The first one is the aforementioned growth of the first hump of the barrier which is the highest one in almost all  nuclei with $N>170$.   The other factor is the lowering of the energy of the oblate minimum which becomes the ground state in the nuclei around $N=178$ (cf. discussion of ground-state deformations in Sec. \ref{ssGSP}). 
The shift of the ground state from the prolate to the oblate minimum, see also Fig.~\ref{fig:spe},  gives an additional increase of the barrier height up to over 1 MeV.
The energy difference between prolate and oblate minima has its largest value in the $N=178$ isotones what, added to  the large first barrier, produces extremely long half-lives. They exceed (in the NRS mode)  $T = 10^7$ s in Cn and $Z=114$   and $T = 10^{14}$ s in $Z=120$.

In the neutron rich Hs isotopes the NRS barriers extend up to relatively large values of the quadrupole moment. Therefore their  transition probabilities are smaller than in the RS mode. This fact together with the long  $\alpha$ decay half-lifes implies that in this region one can find isotopes with very long half-lives.  The RS fission is the dominant decay mode. However these very neutron rich Hs isotopes are extremely difficult to synthesize using contemporary experimental techniques. 

\subsection{The region $ N \ge 180$}

The saddle point of the first hump of the barrier takes its maximal energy value at $N=182$. We can observe in these isotones another very high maximum of the fission half-lives in all elements. In contrast, the $N=180$ isotones decay a few orders of magnitude faster than the neighboring nuclei with $N=178$ and $N=182$.  The influence of the oblate minimum and the high first barrier is not strong enough to enlarge the fission half-lives here. In two isotones with $N=180$, namely Cn and $Z=114$, the NRS fission is the dominant process with half-lives shorter than for the $\alpha$ decay.

The energy difference between the oblate and the prolate minima as well as between the absolute values of their quadrupole moments shrink  in  $N=180$ and $N=182$ continuously.  Finally, one founds that all $N=184$ isotones do have a spherical ground state. This indicates a magic number at  $N=184$. Nevertheless, this feature does not have a big impact  on the fission half-lives and the region of the most stable nuclei is slightly shifted towards the neutron deficient isotopes.

At $N=178$ and $N=182$ the $\alpha$ decay half-lives also reach their maxima although they are less pronounced than for the fission half-lives. They correspond to the minima of $Q_\alpha$ that can be observed for these neutron numbers, see 
Fig.~\ref{fig:Qalpha}.  For the $Z  \ge 116$ elements the  $\alpha$ emission is the fastest decay process for isotopes lighter than $N=184$. Most of them live long enough to allow the synthesis of these nuclei. The extremely long fission half-lives for the $N=178$ and $N=182$ isotopes are larger than  the $\alpha$ decay half-lives in Cn and $Z=114$ whereas for  $N=180$ NRS fission is the dominant mode.

The isotopes with neutron number larger than the magic $N=184$ are characterized by a rapid decrease of the fission half-lives with increasing neutron number. This is a consequence  of the decrease of energy along the whole energy barrier. The second minimum goes down below the ground state and the height of the first barrier reduces substantially when heavier nuclei are considered. These strong trends can not be balanced by the few MeV decrease of the ground state energy due to the octupole deformation. Consequently, for the elements from Ds to $Z=120$, fission becomes the dominant decay mode with half-lives  below $T = 10\; \mu$s. The $\alpha$ decay half-lives do not vary strongly along the isotopic chain in this region. The  fluctuations are associated with the changes of the ground state deformations of the parent and the daughter nuclei. In this region the decrease of the $\alpha$ decay half-lives with the proton number is also visible. Most isotopes in this region can not be synthesized due to the very short fission (in the proton deficient nuclei) or $\alpha$ emission (in the proton rich isotones) half-lives. The experimental limit of $T = 1$ ms is exceeded only for a  few nuclei.

To conclude this section we would like to present in Fig.~\ref{fig:T1/2_main} the shortest half-lives of each isotope in the form of the chart of SHE's. From this figure it is easy to distinguish the regions where each decay mode plays the most important role. The predominant decay mode, specially in the proton rich region, is the  $\alpha$ emission.
Roughly speaking, the spontaneous fission in the RS mode is dominant in the proton deficient nuclei with $Z\le 104$ for $N\ge 158$, $Z\le 108$ for $N\ge 170$, and $Z\le 116$ for $N\ge 186$. In the central part of the diagram the region with the fastest decay in the NRS fission channel is defined by $108\le Z\le 114$ and $170\le N\le 180$. 

Two regions of  long living nuclei can be found also in Fig.~\ref{fig:T1/2_main}. The first one includes nuclei around $^{268}$Sg$_{162}$ where half-lives reach $T = 10^4$ s. The other ``island of stability" is associated with the anomalous long fission half-lives at $N=178$ and $N=182$. At these neutron numbers (and also at $N=184$) long living isotopes of Cn and $Z=114$ elements can be found. The half-lives of these isotopes are also longer than $T = 10^6$ s.
Very long half-lives characterize also the  Hs isotopes decaying through the RS fission.

\section{Summary and conclusions}

In this work the Hatree-Fock-Bogoliubov theory with large basis size and the density dependent Gogny force as interaction have  been used to study the most relevant properties of 160 heavy and super-heavy elements as well as their predominant
decay modes.  In order to keep the calculations as general as possible we allow for wave functions with 
different symmetries, namely axially symmetric and reflexion symmetric, axially symmetric and non reflexion symmetric and
lastly triaxial wave functions. After a  thorough analysis along  different fission paths and considering 
that our calculations are extremely CPU time  demanding due to the  large basis to study 160 nuclei
 we perform most calculations in the axially symmetric approaches.

The ground state deformations $\beta_2, \beta_3, \beta_4$ and  $\beta_6$ as well as $Q_\alpha$ factors,
pairing properties and the two neutron and two proton separation energies are thoroughly discussed. The single particle energies are used as a guide for the interpretation of these properties. 
   
     The fission paths for three representative nuclei, namely $^{274}$Hs,  $^{278}$Ds and $^{282}$Cn, are analyzed in detail.  Properties like mass parameters, pairing features and the variation of the action among others  are calculated along the fission path with wave functions  of the three types  mentioned above. We find that though the shape of the fission barrier has a large impact on the fission half-lives, the mass parameter also plays an important role. In such a way that  wave functions
with larger barriers  and smaller mass parameter may tunnel easier  than alternative ones with smaller barriers and larger mass. Since in general axially symmetric wave functions do have smaller masses than the triaxial ones, the restriction to axial symmetry is a good option to perform a systematic study of the half-lives of SHE.
   In the second part of the paper a thorough study of the shapes of the barriers in the AS-RS and AS-NRS is performed. A reasonable explanation of heights and shapes as well as of octupole effects is obtained on the basis of the single particle energy levels. The two dimensional ($Q_2, Q_3$) energy contour plots  for the nuclei $^{274}$Hs and $^{282}$Cn  allow to disentangle the different fission paths and its possible interconnections. 
  
   In the third part of the paper the half-lives of all studied nuclei are calculated for the different decay modes, namely  $\alpha$-decay and along the different fission paths. We find clear tendencies with the neutron number easily explainable on the basis of the behavior of $Q_\alpha$ factors and barrier shapes, respectively.  In particular, we find that the  $\alpha$ emission is the predominant decay mode, specially in the proton rich region. Concerning the spontaneous fission we obtain that  the RS mode is dominant for the proton deficient nuclei in medium mass SHE's  with $Z\le 104$ for $N\ge 158$, $Z\le 108$ for $N\ge 170$, and in the region of the heaviest SHE's with $Z\le 116$ for $N\ge 186$. The fastest decay in the NRS fission channel takes place for $108\le Z\le 114$ and $170\le N\le 180$. The  long living nuclei can be found two regions. The first one is in the vicinity of $^{268}$Sg$_{162}$ where half-lives around  $T = 10^4$ s are obtained. The anomalous long fission half-lives for proton deficient isotones with  $N=178$, $N=182$ and $N=184$ create the second region. At these neutron numbers several isotopes of the elements Cn and $Z=114$  are found with half-lives  longer than $T = 10^6$ s.
Very long half-lives characterize also the Hs isotopes decaying through the RS fission.
The nuclei beyond $Z=120$ and $N=184$ have half-lives too short to be detected within contemporary experimental limit of $T = 10\; \mu$s.

In conclusion we have presented a systematic study of SHE within the self-consistent HFB with very general wave functions and large configuration space. Our calculations provide an overall interpretation of the systematics and global properties of these elements and its decay modes. In general we find a reasonable good agreement with the experimental data.   


\acknowledgments  
The authors would like to thank Prof.  Krzysztof Pomorski for a  careful reading of the manuscript.
Contribution of Luis M. Robledo to this paper is gratefully acknowledged.
Work partially supported by Grant No. DEC-2011/01/B/ST2/03667 from the National Science Centre (Poland) and by the Spanish Ministerio de Educaci\'on y Ciencia  under contracts  FPA2009-13377-C02-01 and  FPA2011-29854-C04-04 and  the Spanish Consolider-Ingenio 2010 Programme CPAN (CSD2007-00042).



\end{document}